\documentclass[prd,tightenlines,nofootinbib,superscriptaddress]{revtex4}

\usepackage{amsfonts,amssymb,amsthm,bbm,amsmath,wasysym}
\usepackage{hyperref}

\usepackage{color,psfrag}
\usepackage[dvips]{graphicx}

\newcommand{\C}{{\mathbb C}}

\newcommand{\R}{{\mathbb R}}

\newcommand{\cH}{{\mathcal H}}

\newcommand{\cN}{{\mathcal N}}

\newcommand{\cP}{{\mathcal P}}
\newcommand{\cT}{{\mathcal T}}

\newcommand{\cS}{{\mathcal S}}

\newcommand{\SU}{\mathrm{SU}}

\newcommand{\SL}{\mathrm{SL}}
\newcommand{\SO}{\mathrm{SO}}

\newcommand{\SB}{\mathrm{SB}}
\newcommand{\SH}{\mathrm{SH}}
\renewcommand{\H}{\mathrm{H}}

\newcommand{\be}{\begin{equation}}
\newcommand{\ee}{\end{equation}}
\newcommand{\beq}{\begin{eqnarray}}
\newcommand{\eeq}{\end{eqnarray}}
\newcommand{\bes}{\begin{eqnarray}}
\newcommand{\ees}{\end{eqnarray}}

\newcommand{\mat} [2] {\left ( \begin{array}{#1}#2\end{array} \right ) }

\newcommand{\su}{{\mathfrak su}}

\renewcommand{\sl}{{\mathfrak sl}}

\newcommand{\tr}{{\mathrm{Tr}}}
\newcommand{\im}{{\mathrm{Im}}}
\newcommand{\f}{\frac}

\def\nn{\nonumber}
\def\pp{\partial}

\def\ka{\kappa}

\def\eps{\epsilon}

\newcommand{\id}{{\mathbb{I}}}
\def\ka{\kappa}

\def\vN{\vec{N}}
\def\vC{\vec{C}}
\def\bz{\bar{z}}
\def\vsigma{\vec{\sigma}}
\def\hu{\hat{u}}
\def\hv{\hat{v}}

\def\vx{\vec{x}}
\def\arr{\rightarrow}
\def\Om{\Omega}
\def\tH{\widetilde{H}}

\begin{document}

\title{Closure constraints for hyperbolic tetrahedra}

\author{{\bf Christoph Charles}}\email{christoph.charles@ens-lyon.fr}
\affiliation{Laboratoire de Physique, ENS Lyon, CNRS-UMR 5672, 46 all\'ee d'Italie, Lyon 69007, France}

\author{{\bf Etera R. Livine}}\email{etera.livine@ens-lyon.fr}
\affiliation{Laboratoire de Physique, ENS Lyon, CNRS-UMR 5672, 46 all\'ee d'Italie, Lyon 69007, France}

\date{\today}

\begin{abstract}

%

We investigate the generalization of loop gravity's twisted geometries to a q-deformed gauge group. In the standard undeformed case, loop gravity is a formulation of general relativity as a diffeomorphism-invariant $\SU(2)$ gauge theory. Its classical states are graphs provided with algebraic data. In particular closure constraints at every node of the graph ensure their interpretation as twisted geometries. Dual to each node, one has a polyhedron embedded in flat space $\R^3$. One then glues them allowing for both curvature and torsion.
It was recently conjectured that q-deforming the gauge group $\SU(2)$ would allow to account for a non-vanishing cosmological constant $\Lambda\ne 0$, and in particular that deforming the loop gravity phase space with real parameter $q\in\R_{+}$ would lead to a generalization of twisted geometries to a hyperbolic curvature.
Following this insight, we look for generalization of the closure constraints to the hyperbolic case. In particular, we introduce two new closure constraints for hyperbolic tetrahedra. One is compact and expressed in terms of normal rotations (group elements in $\SU(2)$ associated to the triangles) and the second is non-compact and expressed in terms of triangular matrices (group elements in $\SB(2,\C)$). We show that these closure constraints both define a unique dual tetrahedron (up to global translations on the three-dimensional one-sheet hyperboloid) and are thus ultimately equivalent.



\end{abstract}

\maketitle

\section*{Introduction}

We investigate the characterization of a hyperbolic tetrahedron (embedded in a homogeneous hyperbolic space, with constant negative curvature) through four group variables interpreted as the respective ``normals'' to its four triangular faces and related by closure constraints. The main purpose of this study is its application to loop quantum gravity with a non-vanishing cosmological constant and the generalization of its associated twisted geometries to the hyperbolic case.

\begin{figure}[h]
\includegraphics[scale=0.7,angle=-90]{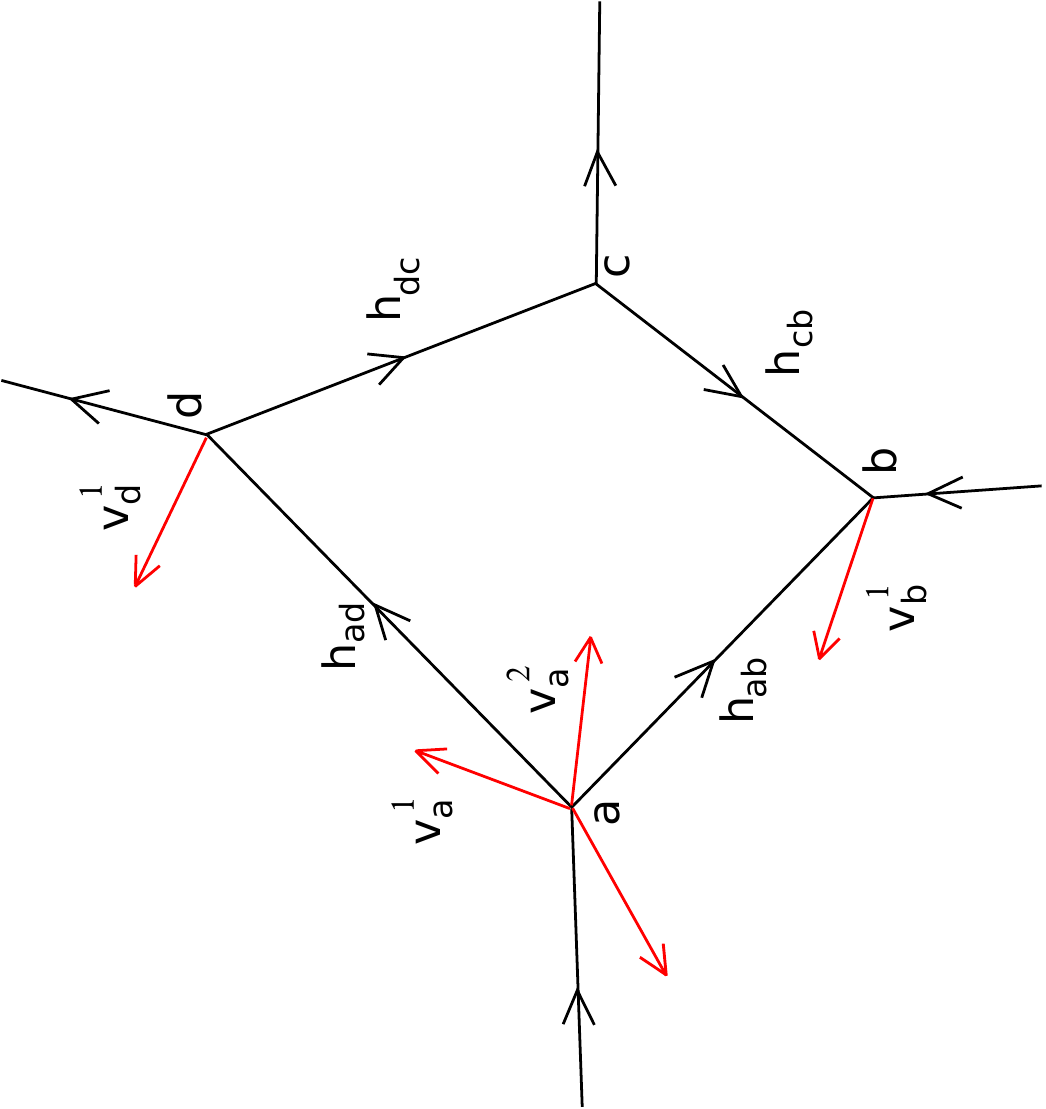}
\caption{\label{fig:GraphEx}The discrete setting of loop quantum gravity on a fixed graph: each edge end carries a spinor in $\mathbb{C}^2$, which defines a vector in $\R^{3}$, and each edge carries a group element in $\SU(2)$ mapping one vector onto the other by a 3d rotation; this can be mapped onto a dual twisted geometry with convex polyhedra dual to each node and glued by area matching constraints through each edge.}
\end{figure}

Indeed, loop quantum gravity offers a non-perturbative framework for a canonical quantization of general relativity (for a review, see  \cite{rovelli2007quantum,thiemann2007modern}). Its quantum states of geometry are the spin network states and can be understood as the quantization of twisted geometries. These are a generalization of Regge geometries allowing for torsion (which encodes extrinsic curvature in the context of loop quantum gravity) introduced by Freidel \& Speziale \cite{twisted1} and then further developed as the natural geometrical interpretation of loop quantum gravity \cite{Livine:2011vk,Livine:2011up,Dupuis:2012yw,Freidel:2013bfa}. They can be understood as  patchwork of flat polyhedra, individually embedded in flat 3d space, glued together through area matching conditions across their triangular faces but without face shape matching. Mathematically, given a fixed (oriented) graph (defined combinatorially), we consider the symplectic reduction $T^*\SU(2){}^{\times E}//\SU(2)^V$  with one copy of $T^*\SU(2)$ on each edge and one gauge invariance imposed at each node or vertex of the graph. The gauge invariance at each vertex is imposed by closure constraints, which in turn ensure the existence of a flat convex polyhedron dual to the vertex, uniquely determined by the $T^*\SU(2)$ elements living on the edges incident to that vertex. These structures are illustrated in fig.\ref{fig:GraphEx}.  

The goal is to study the generalization of this setting to curved polyhedra, individually embedded in a hyperbolic space of given constant curvature. The main purpose is to understand how to account for a non-vanishing cosmological constant in the loop quantum gravity formalism, but it can also be considered as a first step towards understanding the emergence of curvature in the coarse-graining of spin network states.

\medskip

So taking into account the cosmological constant in loop quantum gravity is a central issue. At the classical level,  $\Lambda$  is simply taken into account by a volume term in the Hamiltonian constraint, while it does not affect the local (Lorentz) gauge invariance. Nevertheless, based on the extensive work on 3d quantum gravity, the persistent relation between gravity and Chern-Simons theories, the spinfoam quantization of 4d BF theory and the asymptotics of spinfoam amplitude, it has been conjectured that the cosmological constant $\Lambda$ should also enter the kinematics of the theory at the quantum level through a quantum deformation of the local Lorentz group or of the local $\SU(2)$ gauge group.

In the 3d case, the resulting q-deformation of the corresponding Lorentz group is very well understood from the canonical point of view in the combinatorial quantization framework for Chern-Simons theory \cite{Alekseev:1994pa,Buffenoir:2002tx,cat} and from the path integral perspective both from the Chern-Simons quantization \cite{Witten1} and the Turaev-Viro spinfoam model \cite{Mizoguchi:1991hk} with the asymptotics of the q-deformed $\{6j\}$-symbols \cite{woodward}. From the canonical perspective of loop quantum gravity, which is subtlety different from the Chern-Simons theory, this issue is not entirely settled despite some recent interesting  efforts \cite{KauffmanBracketLQG,pranzetti} and \cite{Dupuis:2013lka,HyperbolicPhaseSpace,Bonzom:2014bua,Dupuis:2014fya}.

For instance, a deformed phase space for loop quantum gravity on fixed graphs has been introduced in \cite{HyperbolicPhaseSpace} by switching $T^*\SU(2)$ for $\SL(2,\C)$ and defining a braided action of $\SU(2)$ at the graph's nodes. In three space-time dimensions, it has been proved that this deformed phase space (for 3-valent graphs) provided with suitable gauge-invariant flatness constraints can be given a geometrical interpretation as equivalent to the data of hyperbolic triangles glued together as a 2d surface embedded in the three-dimensional (upper) hyperboloid $\SO(3,1)/\SO(3)$ of unit time-like vectors in the (3+1)d Minkowski space. The objective is to extend such a construction to the 4d case. 

This requires understanding how the (q-)deformed closure constraints, encoding the (braided) action of the $\SU(2)$ gauge group at the graph's nodes, can be interpreted as ensuring the existence of a hyperbolic  polyhedron dual to the node. In this paper, we focus on the simplest case of a 4-valent node and the definition of the dual hyperbolic tetrahedron.

\medskip

In the standard case, considering a 4-valent node, a vector $\vN_{i}\in\R^3\sim\su(2)$, living in the flat 3d Euclidean space identified to the $\su(2)$ Lie algebra, is associated to each edge attached to the node and labeled by $a=1..4$.  Then we impose the following very simple closure constraints at the node:
\be
\sum_{a=1}^4 \overrightarrow{N}_a = \overrightarrow{0}\,.
\ee
On the one hand, the geometrical interpretation is provided by Minkowski's theorem: there exists a unique (convex) tetrahedron such that the $\vN_{a}$ are the normal vectors of its four triangular faces, that is their norm is the area of the corresponding triangle while their direction is orthogonal to the triangle's plane. This actually applies to an arbitrary number of faces and arbitrary convex polyhedra (see \cite{Bianchi:2010gc,Livine:2013tsa}).
On the other hand, we use the Poisson brackets inherited from the $T^*\SU(2)$ structure
\be
\{N_{a}^{i},N_{a}^{j}\}=\eps^{ijk}N_{a}^{k}\,,
\ee
and the brackets with the closure constraints, or Gauss law in the context of loop gravity, generate the $\SU(2)$ action imposing the gauge invariance at the node.
The challenge  is to extend this construction to hyperbolic tetrahedra.

From the analysis of the deformed phase space introduced in  \cite{HyperbolicPhaseSpace}, based on replacing $T^*\SU(2)$ by $\SL(2,\C)$ and using the Poisson bracket defined by the classical $r$-matrix of $\sl(2,\C)$, vectors $\vN$ of $\R^3$ were replaced by (lower) triangular 2$\times$2 matrices $\ell$:
\be
\ell=\mat{cc}{\lambda & 0 \\ z & \lambda^{-1}}
\in\SB(2,\C)\,,
\ee
endowed with a non-canonical Poisson bracket defined in terms of a deformation parameter $\ka>0$:
\be
\{\lambda,z\}=\f{i\ka}{2}\lambda z,\quad
\{\lambda,\bz\}=-\f{i\ka}{2}\lambda \bz,\quad
\{z,\bz\}={i\ka}(\lambda^{2}-\lambda^{-2}),
\ee
and a mapping to 3-vectors $\vec{T}=(2\ka)^{-1}\,\tr \ell\ell^{\dagger}\vsigma$ by projecting on the Pauli matrices $\sigma_{a}$.
The proposed Gauss law, generating the (braided) $\SU(2)$ action, is then simply $\ell_{1}..\ell_{4}=\id$, where the abelian sum of the standard flat case is now replaced by the non-abelian group product. We recover all the classical setting as $\ka\rightarrow 0$.
This leads to the natural question whether this deformed Gauss law can be interpreted as a closure constraint for hyperbolic tetrahedra.

\smallskip

In this context, our investigation led us to consider two  closure constraints, both rather natural but algebraically quite different:
\begin{itemize}

\item A compact closure constraint $H_{1}..H_{4}=\id$ in terms of $\SU(2)$ group elements $H_{i}\in\SU(2)$:

Each $H_{i}$ associated to a hyperbolic triangle comes naturally when composing the boosts around the triangle embedded in the hyperboloid $\SL(2,\C)/\SU(2)$ and turns out to define a straightforward ``normal rotation'' to the hyperbolic triangle. Indeed the rotation angle gives exactly the triangle area (up to the curvature radius of the hyperboloid) and the axis of rotation is orthogonal to the triangle's plane (suitably transported parallelly).

\item A non-compact closure constraint $L_{1}..L_{4}=\id$ in terms of triangular matrices $L_{i}\in\SB(2,\C)$:

Exploiting that $\SB(2,\C)$ defines a section of the quotient $\SL(2,\C)/\SU(2)$, each $L_{i}$ corresponds to a suitable composition of boosts around two edges of the triangle, similarly to the usual wedge product (or vector product) of two edge vectors in $\R^3$ defining the normal vector of a flat triangle. This proposal seems to fit best with the deformed phase space and  symplectic structure defined in  \cite{HyperbolicPhaseSpace,Dupuis:2014fya}

\end{itemize}

We will prove in this paper how both sets of closure constraint lead to the existence of a unique corresponding hyperbolic tetrahedron. These two closure constraints clearly reflect two dual points of view on the hyperboloid realized as the quotient $\SL(2,\C)/\SU(2)$, the first focusing on the stabilizer group $\SU(2)$ and the second focusing on the coset itself parametrized as a group.

\smallskip

In a first section, we will review in details the case of the flat tetrahedron, the closure constraints and a proof that determine a unique dual tetrahedron.
Section II tackles the algebraic description of a hyperbolic triangle, embedded in the 3d hyperboloid $\SL(2,\C)/\SU(2)$. The third section deals with the analysis of the compact closure constraint while the fourth section investigates the non-compact closure constraint.
We will conclude with a discussion of the advantages and shortcomings of these two closure constraints and their application to loop quantum gravity with a cosmological constant.

\medskip

During the final editing stage of the present short article, we became aware of a new very interesting work \cite{Haggard:2014xoa} by Haggard and al., developed parallely to our own investigation. Although their approach is not the same, there seems to be some non-trivial overlap, especially on the introduction of a compact closure constraint for the hyperbolic tetrahedron. We nevertheless seem to be investigating rather complementary aspects of discrete hyperbolic geometry, and we recommend \cite{Haggard:2014xoa} to the reader interested by the relation between loop quantum gravity with a cosmological constant and $\SL(2,\C)$ Chern-Simons theory.

\section{The flat tetrahedron from normal vectors}

\subsection{Closure constraints}

Let us review the closure constraints in flat case. Indeed we know from Minkowski's theorem (e.g. \cite{Bianchi:2010gc}) that there is a unique way (up to a translation) of reconstructing a polyhedron from the normals to its faces. The situation is even simpler in the case of a tetrahedron, in which case its reconstruction from the normals is rather straightforward.

\begin{figure}[h!]
\includegraphics[scale=0.8, angle=-90]{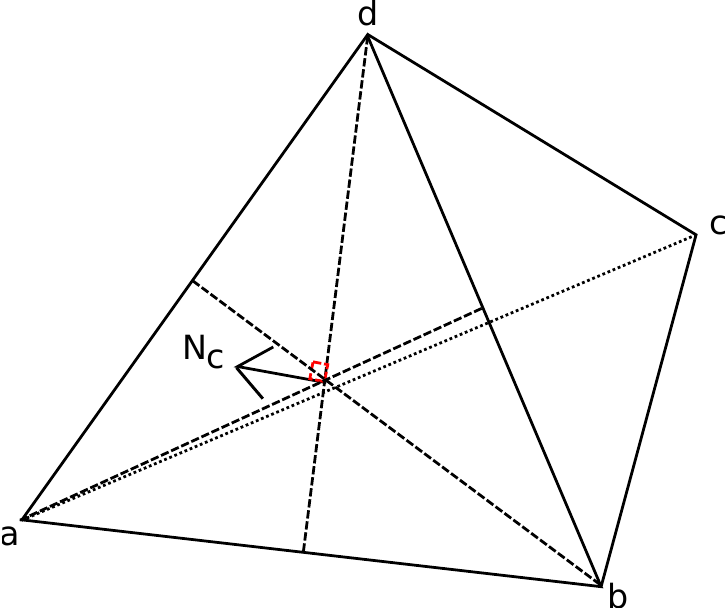}
\caption{A flat tetrahedron, with normals are labeled by the opposite vertex.}
\label{fig:FlatTetra}
\end{figure}

So let us study the standard flat tetrahedron (fig.\ref{fig:FlatTetra}) in $\R^3$ before generalizing to the hyperbolic case in the next sections. We define a flat tetrahedron by its four vertices $(a,b,c,d)$ with the triangle and their normals labeled by their respective opposite vertex. The game is to find a relation satisfied by these normals and such that once assumed, it is always possible to reconstruct a unique corresponding tetrahedron.

Using the normals, we only study the tetrahedron up to a global translation. As a consequence, all the information needed to characterize the tetrahedron is in the vectors $\overrightarrow{ab}$, $\overrightarrow{ac}$, $\overrightarrow{ad}$, defining the relative position of its four vertices.

Let us assume that the tetrahedron is not degenerate, i.e the four points do not lay in the same plane. We further assume that the vectors $(\overrightarrow{ab}$, $\overrightarrow{ac}$, $\overrightarrow{ad})$ are oriented such that $(\overrightarrow{ab}\times \overrightarrow{ac})\cdot\overrightarrow{ad}>0$, else we would simply switch the points $b$ and $c$.

We define the outward vectors normal to the faces with their norm given by the face area:
\begin{eqnarray}
\overrightarrow{N}_b &=& \frac{1}{2} \overrightarrow{ad} \times \overrightarrow{ac}\,, \\
\overrightarrow{N}_c &=& \frac{1}{2} \overrightarrow{ab} \times \overrightarrow{ad}\,, \\
\overrightarrow{N}_d &=& \frac{1}{2} \overrightarrow{ac} \times \overrightarrow{ab}\,,
\end{eqnarray}
and:
\begin{equation}
\overrightarrow{N}_a = \frac{1}{2} \overrightarrow{bc} \times \overrightarrow{bd} = \frac{1}{2} \left(\overrightarrow{ad} \times \overrightarrow{ab} + \overrightarrow{ab} \times \overrightarrow{ac} - \overrightarrow{ad} \times \overrightarrow{ac} \right)\,.
\end{equation}

These normals obviously satisfy the symmetric relation:
\begin{equation}
\overrightarrow{N}_a + \overrightarrow{N}_b + \overrightarrow{N}_c + \overrightarrow{N}_d = 0
\end{equation}
This is the closure constraints for the flat tetrahedron.

\subsection{Reconstructing the tetrahedron}

It is the reciprocal relation that's particularly interesting for us: if we start with a set of four vectors satisfied such closure constraints, thus summing to 0, there exists a unique tetrahedron (up to translation) such that these vectors are its face normals.
Let us have a closer look at the reconstruction procedure of this tetrahedron.

Starting with four vectors satisfying $\vC=\vN_{a}+\vN_{b}+\vN_{c}+\vN_{d}=0$, we want them to  represent the normals to the tetrahedron faces. Thus each of them defines an orthogonal plane in which the corresponding triangle lives. The tetrahedron edges are then at the intersection of two of those planes. This determines the tetrahedron up a global scale, which is given by computing the tetrahedron volume from the mixed product of the normal vectors.

In more details, let us first consider the case that $U\equiv \vN_{d}\cdot(\vN_{c}\times \vN_{b})>0$. If $U=0$, the tetrahedron would be degenerate and we discard this case. If $U<0$, we would simply follow the same  reconstruction procedure as below by after switching the role of the normal vectors $\vN_{b}$ and $\vN_{c}$.

Let us assume that there exist a tetrahedron corresponding to these normal vectors, defined as above with the three the vectors $(\overrightarrow{ab}$, $\overrightarrow{ac}$, $\overrightarrow{ad})$ positively-oriented . Then the edge shared by two faces would be orthogonal to the two corresponding normal vectors. So tacking their cross-product, we would get the edge vector up to normalization. We compute explicitly the cross-product of two normal vectors in order to see that the correct coefficient is the tetrahedron volume:
\be
\overrightarrow{N}_b \times \overrightarrow{N}_c
\,=\, \frac{1}{4}\left(\overrightarrow{ad} \times \overrightarrow{ac} \right) \times \left(\overrightarrow{ab} \times \overrightarrow{ad} \right) \\
\,=\, \frac{1}{4} \overrightarrow{ad} \cdot \left(\overrightarrow{ab} \times \overrightarrow{ac} \right) \overrightarrow{ad} \\
\,=\, \frac{3}{2} V \overrightarrow{ad}\,,
\ee
where $V=(\overrightarrow{ab}\times \overrightarrow{ac})\cdot\overrightarrow{ad}/6$ is the oriented volume of the tetrahedron. We have chosen $V>0$ by definition.

To finish to extract the edge vectors, we need an independent formula to compute the volume.
We extract it from the triple product of the normal vectors, indeed:
\be
U\,=\,
\overrightarrow{N}_d \cdot \left(\overrightarrow{N}_c \times \overrightarrow{N}_b \right)
\,=\, \frac{1}{2} \left(\overrightarrow{ab} \times \overrightarrow{ac} \right) \cdot \left(\frac{3V}{2} \overrightarrow{ad} \right)
\,=\, \frac{9}{2} V^2>0\,.
\ee
Finally this allows to express the edge vectors explicitly in terms of the normal vectors:
\begin{equation}
\overrightarrow{ad} = \sqrt{\frac{2}{U}} \overrightarrow{N}_b \times \overrightarrow{N}_c,\quad
\overrightarrow{ab} = \sqrt{\frac{2}{U}} \overrightarrow{N}_c \times \overrightarrow{N}_d,\quad
\overrightarrow{ac} = \sqrt{\frac{2}{U} } \overrightarrow{N}_d \times \overrightarrow{N}_d,\quad
U=\vN_{d}\cdot(\vN_{c}\times \vN_{b})>0\,.
\end{equation}
This is only valid, as long as the volume is not zero, that is as long as the tetrahedron is not degenerate. As said above, if $U<0$, we switch the role of $\vN_{b}$ and $\vN_{c}$ in the reconstruction process.

To conclude this analysis, reciprocally, it is now easy to check that the tetrahedron defined by those edge vectors as before does indeed have the correct expected normal vectors.

\subsection{The  flat tetrahedron phase space}

In order to study the dynamic of the tetrahedron in the context of quantum geometry and quantum gravity, we have to embed it in a phase space with a symplectic structure. It turns out that there is a natural Poisson structure inherited from the $\mathfrak{su}(2)$ Lie algebra structure, see e.g. \cite{kapovich1996} and more recently \cite{Bianchi:2010gc,Livine:2013tsa,Haggard:2014gca}. Then gluing together these tetrahedron phase space through area matching constraints and $\SU(2)$ group elements encoding the parallel transport, we can the phase space of twisted geometries on a field graph for loop quantum gravity.

%

Thus equipping the space of normal vectors with the Poisson bracket on $\mathbb{R}^3\sim\su(2)$, that is $\{x_i, x_j \} = \epsilon_{ijk} x_k$ for a 3d vector $x$, we define:
\begin{equation}
\label{bracketN}
\{ N^i_a, N^j_b \} = \delta_{ab} \epsilon^{ij}_k N^k_a
\end{equation}
where $N^i_a$ is the $i^\textrm{th}$ component of the normal vector of face $a$. The closure condition $\vC=\sum_a \overrightarrow{N}_a = \overrightarrow{0}$ turns out to be a first-class closure constraint. It generates $\SO(3)$ global rotations of the four normal vectors.

However so far, we  have a degenerate Poisson structure. It is possible to remedy this. Let us write $N_a$ for the norm of $\overrightarrow{N_a}$. It commutes with all the vector components, $\{N_a, N_a^i\} = 0$, and we identify it as the degenerate sector of our space. As a consequence, we consider the space of sets of four vectors with fixed norms. Intuitively, this corresponds to the space of tetrahedra with fixed triangle areas but that are not necessarily closed. Upon performing the symplectic reduction (double quotient) by the closure constraints $\vC=0$, we obtain a space $\left(\mathcal{S}^2\right)^4 // \vC$ isomorphic to $\mathcal{S}^2$ equipped with a symplectic structure \cite{kapovich1996,Freidel:2009nu}. This is the Kapovich-Millson phase space of (shape of) tetrahedra with fixed triangle areas (up to translations and rotations).

There is  a more recent approach using spinors \cite{twisted1,spinor-lqg} that allows us to define a symplectic space without fixing areas, and allowing to naturally glue the tetrahedron phase spaces into a twisted geometry phase space on a given graph, see e.g. \cite{Livine:2011up}.
So, we consider spinors, that is two-dimensional complex vectors $| z_a \rangle\in\C^{2}$ living in the fundamental representation of $\SU(2)$.
We take the canonical symplectic structure:
\begin{equation}
\{ z_a^A, \overline{z}_b^B \} = i \delta_{ab} \delta^{AB}\,,
\end{equation}
where $A,B=0,1$ are spinor indices. This Poisson structure can be derived from the following action:
\begin{equation}
S = -i \int dt\,\sum_a \langle z_a | \partial_t z_a \rangle .
\end{equation}
We define the 3-vectors by projecting the spinors onto the Pauli matrices, $N^i_a = \langle z_a | \sigma^i | z_a \rangle$, which leads to the exact same Poisson bracket as before in \eqref{bracketN}.

The action $S$ is clearly invariant under  global $SU(2)$ transformations on the spinors $| z_a \rangle$, which correspond to $3d$ rotations on the vectors $\vN_{a}$:
\begin{eqnarray*}
| z_a \rangle &\rightarrow& g| z_a \rangle \\
\langle z_a | &\rightarrow& \langle z_a | g^\dagger
\end{eqnarray*}
More precisely, for a possibly time-dependent transformation, $g(t)\in\SU(2)$, we compute:
\be
S
\,\longrightarrow\,
-i \int dt \,\sum_a \langle z_a | g^\dagger \partial_t \left(g | z_a \rangle\right) 
\,=\,
S -\frac{i}{4} \int dt\, \sum_a  \langle z_a | g^\dagger \partial_t g| z_a \rangle\,.
\ee
Since $g^\dagger \partial_t g\in\su(2)$\footnotemark, the closure constraints $\sum_{a}N^i_a =\sum_{a} \langle z_a | \sigma^i | z_a \rangle=0$ ensure that the extra-term vanishes.
\footnotetext{Explicitly parametrizing the group element as $g=\cos\theta+i\sin\theta \hu\cdot\vsigma$, we have
$g^{-1}\pp g=-i\,\vsigma\cdot\,[(\cos^{2}\theta-\sin^{2}\theta)\pp\theta\,\hu+\cos\theta\sin\theta\pp\hu]$.
}
%
In order to understand the actual content of the phase space, let us gauge-fix this $\SU(2)$-invariance and study the remaining degrees of freedom.

\begin{figure}[h!]
\includegraphics[scale=0.8]{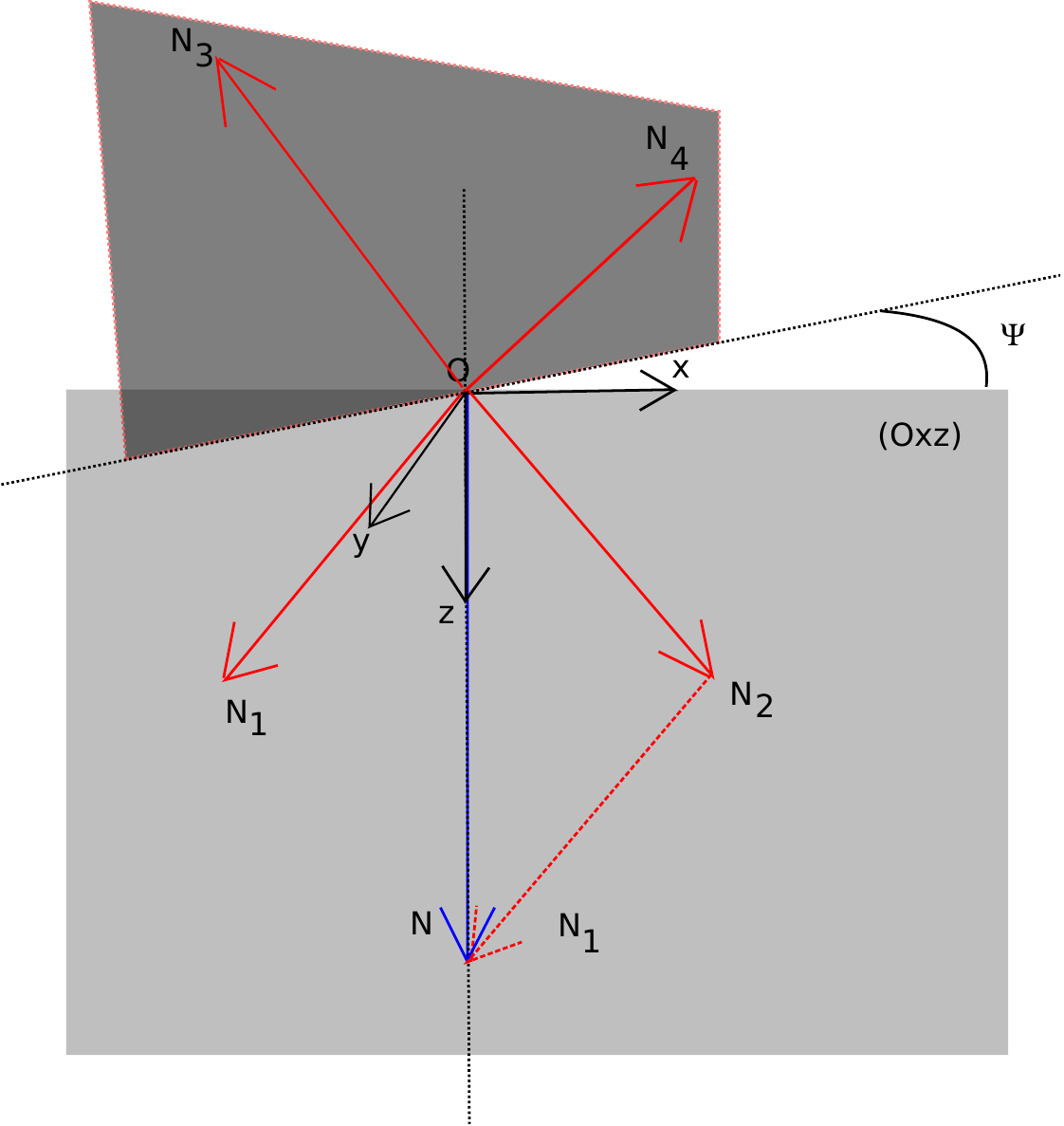}
\caption{Gauge fixing procedure}
\label{fig:GaugeFix}
\end{figure}

As illustrate in fig.\ref{fig:GaugeFix}, a natural gauge fixing choice (which is always possible as long as the tetrahedron is not degenerate) is to align the sum of the first two normals $\vN\equiv\overrightarrow{N}_1 + \overrightarrow{N}_2$ along the $(Oz)$ (half-)axis and to fix the two normals, $\vN_{1}$ and $\vN_{2}$ to lay in the $(Oxz)$ plane.
As a consequence, we can write:
\begin{equation}
| z_1 \rangle = e^{i\phi_1} \begin{pmatrix}
\sqrt{N_1} \cos \frac{\theta_1}{2} \\
\sqrt{N_1} \sin \frac{\theta_1}{2} 
\end{pmatrix}
\end{equation}
and:
\begin{equation}
| z_2 \rangle = e^{i\phi_2} \begin{pmatrix}
\sqrt{N_2} \cos \frac{\theta_2}{2} \\
\sqrt{N_2} \sin \frac{\theta_2}{2} 
\end{pmatrix}
\end{equation}
where $N_1$ and $N_2$ are the norms of the normals, $\theta_1$ and $\theta_2$ are the angles of the normals with respect to $(Oz)$ and $\phi_1$ and $\phi_2$ are possible phases carried by the spinors.
%

For the other couple of spinors, with the sum $\overrightarrow{N}_3 + \overrightarrow{N}_4 = -\overrightarrow{N}$, we can parametrize them similarly, although now they do not necessarily lay in the plane $(Oxz)$:
\begin{equation}
| z_3 \rangle = e^{i\phi_3} \begin{pmatrix}
\sqrt{N_3} \cos \frac{\theta_3}{2} e^{-i\psi} \\
\sqrt{N_3} \sin \frac{\theta_3}{2} e^{i\psi}
\end{pmatrix}
\end{equation}
and:
\begin{equation}
| z_4 \rangle = e^{i\phi_4} \begin{pmatrix}
\sqrt{N_4} \cos \frac{\theta_3}{2} e^{-i\psi} \\
\sqrt{N_4} \sin \frac{\theta_3}{2} e^{i\psi}
\end{pmatrix}
\end{equation}
where $\psi$ is the angle between the plane $(Oxz)$ defined $\overrightarrow{N}_1$ and $\overrightarrow{N}_2$ and the one defined by $\overrightarrow{N}_3$ and $\overrightarrow{N}_4$.


Explicitly computing the action $S$ on this gauge-fixed parametrization of the spinors and ignoring the total derivative (not contributing to the symplectic structure), we get:
\begin{equation}
S = -i \int dt\,\left(i N_1 \partial_t \phi_1 + i N_2 \partial_t \phi_2 + i N_3 \partial_t \phi_3 + i N_4 \partial_t \phi_4 + i N \partial_t \psi \right)\,,
\end{equation}
where $N=N_1 \cos \theta_1 + N_2 \cos \theta_2=N_3 \cos \theta_3 + N_4 \cos \theta_4$ is the norm of the vectors $\vN=\vN_{1}+\vN_{2}=-(\vN_{3}+\vN_{4})$, identified geometrically as the area of the internal parallelogram whose summit are the midpoints of the edges $(13),(23),(24),(14)$ (defined by the two triangles sharing them).

This provides us with a symplectic structure, that we can use for Hamiltonian dynamics, with conjugate angles to the triangle areas and the pair of conjugate variables $\{N,\psi \} = 1$ parametrizing the Kapovitch-Millson phase space.

\section{The hyperbolic triangle}

Let us now turn to the hyperbolic case. We will focus on the geometric and algebraic definition of the face normals, the derivation of suitable closure constraints and the reconstruction procedure of the hyperbolic tetrahedron. We postpone the detailed analysis of the phase space structure to future work.

In this section, we characterize the hyperbolic triangle, as embedded in the 3-dimensional hyperboloid $\SL(2,\C)/\SU(2)$. We will then introduce and discuss the closure constraints in sections III and IV.

\subsection{The 3-hyperboloid: boosts and triangular matrices}

Our curved tetrahedron will  live in the 3d one-sheet hyperboloid defined as the submanifold of $\mathbb{R}^{3,1}$ of timelike vectors satisfying:
\begin{equation}
t^2 - x^2 - y^2 - z^2 = \kappa^2,\qquad t > 0
\end{equation}
where $\kappa$ is the curvature radius of the hyperboloid.
A useful representation of this hyperboloid is given by the isomorphism as vector spaces between $\R^{3,1}$ and the space of Hermitian 2$\times$2 matrices  $\H_{2}$. The identification is given by:
\begin{equation}
\begin{pmatrix}
t \\
x \\
y \\
z
\end{pmatrix} \rightarrow \begin{pmatrix}
t + z & x -i y \\
x +i y & t - z
\end{pmatrix} = M
\end{equation}
This way, the (pseudo-)norm in $\R^{3,1}$ matches the determinant in $\H_2$ for $\det M = (t+z)(t-z) - (x+iy)(x-iy) = t^2 -z^2 -x^2 - y^2$.
In the following, we will take $\kappa=1$ for the sake of simplicity and  we will thus identify $\SH_2$ (the space of Hermitian matrices with determinant $1$) and our hyperboloid $\mathcal{H}_3$.

The action of $\SO(3,1)$ on 4-vectors $(t,x,y,z)$ translates into the action by conjugation of $\SL(2,\C)$ on Hermitian matrices in $\SH_2$. Just the same way that we can explore the whole hyperboloid by acting with a boost in $\SO(3,1)$ on the unit time-like vector $\Om=(1,0,0,0)$, we can act with $\SL(2,\C)$ on the identity matrix and get all points on $\mathcal{H}_3$ by:
\begin{equation}
M = \Lambda \Lambda^\dagger,\qquad \Lambda\in\SL(2,\C)\,.
\end{equation}
This realizes the identification of the homogeneous space $\mathcal{H}_3$ as a coset of the Lorentz group:
\begin{equation}
\mathcal{H}_3 \simeq \SO(3,1) / \SO(3) \simeq \SL(2,\C) / \SU(2)\,.
\end{equation}

Different choices of section of this coset relate to different decompositions of the Lorentz group. We distinguish in particular:
\begin{itemize}
\item The Cartan decomposition:
\begin{equation}
\forall g \in \SL(2,\C),~\exists ! (b,h) \in \SH^{+}_2 \times \SU(2),~ g = bh\,,
\end{equation}
with
$$
b=e^{\f\eta2\hu\cdot\vsigma}=\cosh\f\eta2\id+\sinh\f\eta2\hu\cdot\vsigma,\quad \eta\ge0,\,\hu\in\cS_{2},
\qquad
h=e^{i\f\theta2\hv\cdot\vsigma}=\cos\f\theta2\id+i\sin\f\theta2\hv\cdot\vsigma,\quad\theta\in[0,2\pi],\,\hv\in\cS_{2}\,,
$$
where switching the sign of $\eta$ or $\theta$ would be simply re-absorbed by taking the opposite boost/rotation direction, $-\hu$ or $-\hv$, on the 2-sphere $\cS_{2}$.
In this case, we represent the points $M$ on the hyperboloid by pure boosts in $\SH^{+}(2,\mathbb{C})$  given as positive Hermitian matrices with unit determinant, that is $M=gg^{\dagger}=bb^{\dagger}=b^{2}$.

\item The Iwasawa decomposition:
\begin{equation}
\forall g \in \SL(2,\C),~\exists ! (\ell,h) \in \SB(2,\C) \times \SU(2),~ g = \ell h\,,
\end{equation}
with
$$
\ell=\mat{cc}{\lambda & 0 \\ z & \lambda^{-1}},\quad\lambda>0,\,z\in\C\,.
$$
In this  case, points $M$ on the hyperboloid are represented by lower triangular matrices with real positive diagonal in $\SB(2,\mathbb{C})$  (the special Borel subgroup of the linear group): $M = g g^\dagger = \ell \ell^\dagger$. The advantage of this representation is that $\SB(2,\mathbb{C})$ is indeed a group  unlike the set of pure boosts  $\SH^{+}(2,\mathbb{C})$.
\end{itemize}

Let us point out that we can switch from one representation to the other by writing the Cartan decomposition of a triangular matrix  $\ell$ (see appendix \ref{app:relation} for more details):
\begin{equation}
\ell = bh,\qquad\textrm{with}\quad
h = \frac{\ell + (\ell^{-1})^\dagger}{\sqrt{2 + \mathrm{Tr} \left(\ell \ell^\dagger\right)}},\qquad
b = \frac{1 + \ell \ell^\dagger}{\sqrt{2 + \mathrm{Tr} \left(\ell\ell^\dagger\right)}}\,.
\end{equation}
A point $M=\Lambda\Lambda^{\dagger}$on the hyperboloid , defined by $\Lambda\in\SL(2,\C)$,  is the equivalence class of all group elements $\Lambda h$ with $h\in\SU(2)$. We can choose various representative of this equivalence class, for instance either the pure boost or triangular matrix coming from the Cartan or Iwasawa decomposition of $\Lambda$.
This identifies the 3-hyperboloid $\cH_{3}$ to two natural sections of the coset $\SL(2,\C) / \SU(2)\sim\SH^{+}(2,\mathbb{C})\sim\SB(2,\mathbb{C})$.
The  Lorentz group $\SL(2,\C)$ naturally acts on the hyperboloid by left multiplication and a Lorentz transformation $g\in\SL(2,\C)$ maps:
$$
\Lambda
\overset{g\in\SL(2,\C)}{\longrightarrow}
g\Lambda\,,
\qquad
M=\Lambda\Lambda^{\dagger} 
\overset{g}{\longrightarrow}
g M g^{\dagger}\,.
$$
Now we would like to distinguish 3d rotations from 3d translations on the 3-hyperboloid. Rotations are straightforwardly identified to the action of $\SU(2)$ group elements, that is a rotation $H\in\SU(2)$ acts as $\Lambda \arr H\Lambda$, $M\arr HMH^{\dagger}$. In the pure boost representation, this amounts to the action by conjugation by $H\in\SU(2)$:
\be
\Lambda
\overset{H}{\longrightarrow}
H\,\Lambda
\,=\,
H \, bh
\,=\,
(HbH^{-1})\,(Hh)\,,
\qquad\textrm{with}
\left|
\begin{array}{lcl}
b&\arr&HbH^{-1}\\
h&\arr&Hh
\end{array}
\right.
\ee
The action of rotations on the Iwasawa decomposition is more subtle since for $H\in\SU(2)$, the matrix $H \ell H^\dagger$ is no longer triangular.
The action by conjugation of $\SU(2)$ becomes non-trivial   and we have to adapt the transformation law for $\ell$ :
\begin{equation}
\Lambda
\overset{H}{\longrightarrow}
H\,\Lambda
\,=\,
H \, \ell h
\,=\,
(H\ell H'^{-1})\,(H'h)\,,
\qquad\textrm{with}\quad
\left|
\begin{array}{lcl}
\ell&\arr&H\ell H'^{-1}\\
h&\arr&H'h
\end{array}
\right.\,,
\end{equation}
where the new $\SU(2)$ group element  $H'$ is unique and depends on $H$ but also on $\ell$ . The interested reader will find more details in\cite{HyperbolicPhaseSpace}.

\medskip

Translations on hyperboloid are more tricky than rotations. Since they are identified to the coset $\SL(2,\C)/\SU(2)$, there is an ambiguity in the strict definition of a translation. Different choices of sections of the coset leads to different definitions of translations, which differ by some 3d rotations. The standard choice in special relativity is to use pure boosts $B\in\SH^{+}_{2}$ to represent translations on the hyperboloid, $\Lambda\rightarrow B \Lambda$:
\be
\Lambda
\overset{B}{\longrightarrow}
B\,\Lambda
\,=\,
B \, bh
\,=\,
(Bbk^{-1})\,(kh)\,,
\qquad\textrm{with}\quad
\left|
\begin{array}{lcl}
b&\arr&Bbk^{-1}\\
h&\arr&kh
\end{array}
\right.\,,
\ee
where we need an extra $\SU(2)$ rotation $k$  since the product of two boosts  $Bb$ is not generically Hermitian and needs to be corrected into a pure boost. This leads to the  Thomas precession \cite{Rhodes:2003id} in special relativity.

On the other hand, we can use the Iwasawa decomposition. Translations on the hyperboloid are then given by Borel elements, which now form a group, and a triangular matrix $L\in\SB(2,\C)$ acts as:
\be
\Lambda
\overset{L}{\longrightarrow}
L\,\Lambda
\,=\,
L \, \ell h
\,=\,
(L\ell)\,h\,,
\qquad\textrm{with}
\left|
\begin{array}{lcl}
L\ell&\arr&L\ell\\
h&\arr&h
\end{array}
\right.\,.
\ee
Thus, since it is far more convenient to use translations which form a group, we will identify in this paper translations on the hyperboloid as Borel subgroup elements in $\SB(2,\C)$. And, it will always be simpler to use the Cartan decomposition when analyzing the action of rotations, while it is simpler to use the Iwasawa decomposition when looking at the action of translations.

\medskip

Now that the basic notations are settled, we can focus on the composition of two translations to form a hyperbolic triangle.

\subsection{The triangle from tangent boosts}

We now consider a triangle on $\mathcal{H}_3$.
Studying the closure constraints defining the hyperbolic triangle is a first step towards developing the closure constraints for the hyperbolic tetrahedron. In particular, it will lead us to introduce the $\SU(2)$ holonomy around the triangle, which will play the role of a group-valued normal vector to the triangle.
%

\begin{figure}[h!]
\includegraphics[scale=0.8,angle=-90]{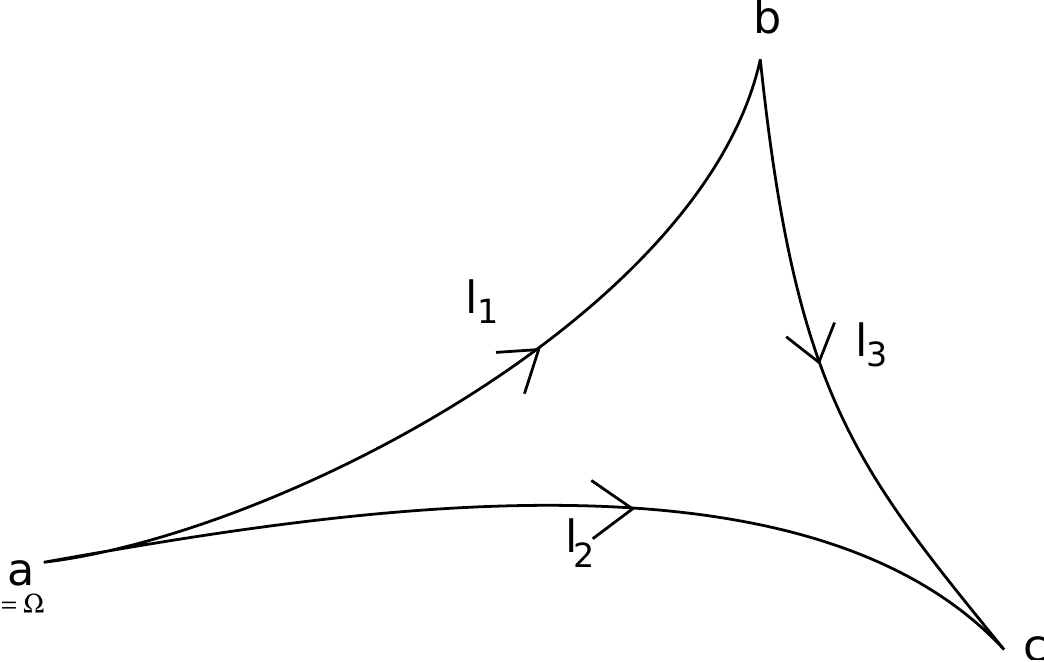}
\caption{Our hyperbolic triangle with labels: we write $a,b,c$ for the three vertices and $1,2,3$ for the three oriented edges, and we set for the sake of simplicity the first point to the hyperboloid origin, $a=\Omega$, $M_{a}=\id$.}
\label{fig:HyperTriangle}
\end{figure}

The triangle is defined by three points (fig. \ref{fig:HyperTriangle}) $a,b,c$.
They are defined by three translations from the hyperboloid origin $\Om=(1,0,0,0)$, namely $\ell_{a,b,c}$. The resulting points are $M_{a}=\ell_{a}\ell_{a}^{\dagger}$ and similarly from $b$ and $c$.
We introduce the three oriented edges of our triangle as $1=(ab)$, $2=(ac)$ and $3=(bc)$. We define the translations along each edge:
\be
\ell_{1}\equiv\ell_{a}^{-1}\ell_{b},\qquad
\ell_{2}\equiv\ell_{a}^{-1}\ell_{c},\qquad
\ell_{3}\equiv\ell_{b}^{-1}\ell_{c}\,.
\ee
These three group elements obviously satisfy the following closure relation:
\begin{equation}
\boxed{\ell_1 \ell_3 \ell_2^{-1} = \id\,.}
\end{equation}

Let us explain its geometrical meaning.
To simplify the discussion, we start with  placing one of these points at the hyperboloid's origin $a=\Om$ or equivalently at the identity $\ell_{a}=\id$, $M_{a}=\id$. Then the Borel group element along the two edges connected to $a$ simplify to $\ell_{1}=\ell_{b}$ and $\ell_{2}=\ell_{c}$, while the group element on the third edge does not simplify $\ell_{3}=\ell_{b}^{-1}\ell_{c}$. Thus starting with the two points $\ell_{b}$ and $\ell_{c}$, we would like to interpret the third group element $\ell_{3}=\ell_{b}^{-1}\ell_{c}$ as the hyperbolic equivalent of the vector from point $b$ to point $c$. Since $\SB(2,\C)$ is not abelian, we have two possibilities:
\begin{itemize}
\item The left composition transformation $m$ that maps one onto the other, $\ell_c =m\ell_{b}$. This is the natural action of a translation  on the 3-hyperboloid $\mathcal{H}_3$ as we defined earlier. And $m$ represents the boost sending the point $M_{b}$ on the point $M_{c}$. Indeed, we have:
$$
mM_{b}m^{\dagger}=m\ell_b\ell_b^{\dagger}m^{\dagger}=\ell_c\ell_c^{\dagger}=M_{c}\,.
$$
\item The right composition transformation $\ell$ that maps one onto the other, $\ell_c = \ell_b\ell$. This group element $\ell$ is the translation that should be done to reach point $c$ if point $b$ was taken as the origin on the hyperboloid. This is actually the proper transformation to consider here. Moreover, the hyperbolic distance between the points $M_{b}= \ell_b \ell_b^\dagger$ and $M_{c}= \ell_c \ell_c^\dagger$ is invariant under Lorentz transformations and is equal to the distance between $\id=\ell_b ^{-1} M_{b} \ell_b^{-1\dagger}$ and $M=\ell_b^{-1} M_{c} \ell_b^{-1\dagger}=\ell\ell^{\dagger}$. That is, the distance between points $b$ and $c$ is given by the argument $\eta$ of the hyperbolic cosine in $\mathrm{Tr} \ell\ell^{\dagger} = 2\cosh \eta$. 

\end{itemize}

We insist on this moot point: the hyperbolic distance between points $b$ and $c$ is given by the (Hermitian) norm of the Borel group element $\ell=\ell_b^{-1}\ell_{c}$ and not by the more obvious translation $m=\ell_{c}\ell_b^{-1}$.
Therefore we do have  a closure relation between Borel group elements attached to each triangle edge, $\ell_1 \ell_3 \ell_2^{-1} = \id$, where the hyperbolic length of every edge is given by the norm of the corresponding group element $\ell_{1,2,3}$.

Now, there is a one-to-one correspondence between sets of three elements $\ell_1$, $\ell_2$, $\ell_3$ satisfying this closure relation $\ell_1 \ell_3 \ell_2^{-1} = \id$ and hyperbolic triangles up to translations with edges defined by these three group elements and vertices given by $\ell_{a}=L\in\SB(2,\C)$ arbitarily chosen on the hyperboloid and $\ell_{b}=L\ell_{1}=\ell_{a}\ell_{1}$ and $\ell_{c}=L\ell_{2}=\ell_{a}\ell_{2}$ derived by translating the first point by the group elements $\ell_{1}$ and $\ell_{2}$. In particular, there is a unique triangle   determined by $\ell_1$, $\ell_2$, $\ell_3=\ell_{1}^{-1}\ell_{2}$ and with its first vertex at the hyperboloid origin $\Om$.


\medskip

We can translate this closure relation into the pure boost parametrization of the hyperboloid.  Let's start from our triangle defined by $\ell_1$, $\ell_2$ and $\ell_3$. For each of these group elements, there is a unique Cartan decomposition $\ell_i = b_i h_i$. So we have:
\be
\ell_1 \ell_3 \ell_2^{-1} \,=\,\id
\quad \Leftrightarrow \quad 
b_1 h_1 b_3 h_3 \,=\,b_2 h_2
\quad \Leftrightarrow \quad 
(b_1) (h_1 b_3 h_1^{-1})  \,=\, b_2 (h_2 h_3^{-1} h_1^{-1})\,.
\ee
So we define the following pure boosts $B_{1,2,3}$ attached to the three edges and a $\SU(2)$ group element $H$ attached to the whole triangle:
\begin{eqnarray}
B_1 &=& b_1 \nn\\
B_2 &=& b_2 \nn\\
B_3 &=& h_1 b_3 h_1^{-1} \nn\\
H &=& h_2 h_3^{-1} h_1^{-1}\,,
\end{eqnarray}
which satisfy the following closure relation:
\begin{equation}
\boxed{B_1 B_3 = B_2 H\,.}
\end{equation}

Let us now consider the converse and start with three pure boosts $B_1$, $B_2$, $B_3$ such that $B_2^{-1} B_1 B_3 \in \SU(2)$.
We would like to show that this uniquely determines the triangle, i.e that we can re-build $\ell_{1,2,3}$ ,and thus also $H$, from these boosts.
We straightforwardly use the Iwasawa decomposition of the boosts $B = \ell h^{-1}$ and define in order $\ell_{1},h_{1}$ and $\ell_{2},h_{2}$ and then $\ell_{3},h_{3}$:
\begin{eqnarray}
B_1 &=& \ell_1 h_1^{-1} \,,\nn\\
B_2 &=& \ell_2 h_2^{-1} \,,\nn\\
h_1^{-1} B_3 h_1 &=& \ell_3 h_3^{-1}\,.
\end{eqnarray}
Although the third edge (opposite to our first vertex $a$) has a slightly special role with the conjugation by non-trivial $\SU(2)$ group element $h_{1}$, it does not cause any problem or block the reconstruction of the triangle.
From here, we compute:
\begin{eqnarray*}
B_2^{-1} B_1 B_3 &=& h_2 \ell_2^{-1} \ell_1 h_1^{-1} \left(h_1 \ell_3 h_3^{-1} h_1^{-1}\right) \\
&=& h_2 \ell_2^{-1} \ell_1 \ell_3 h_3^{-1} h_1^{-1} 
\end{eqnarray*}
For this product to be in $\SU(2)$ as initially assumed, we must have $L = \ell_2^{-1}\ell_1 \ell_3  \in \SU(2)$. But as $L$ is in the $\SB(2,\C)$ subgroup, this implies $L = \id$. Thus we can conclude that as expected:
\begin{equation}
\ell_3 = \ell_1^{-1} \ell_2\,,
\end{equation}
and we recover the closure constraint for the hyperbolic triangle written in terms of triangular matrices.

\medskip

\subsection{The normal rotation}

We have just constructed a holonomy $H\in\SU(2)$ as the product of the group elements $h_{1,2,3}$ around the triangle. This will actually play the role of the normal to the hyperbolic triangle.
Indeed, it has three main characteristics:
\begin{enumerate}
\item It is invariant under translation on the hyperboloid.
\item The rotation axis of $H$ is truly the  normal vector to the triangle (at the origin).
\item Its rotation angle is the deficit angle of the triangle which, in hyperbolic geometry, is the area of the triangle up a $\kappa^2$ factor (that we have set to $1$ for simplicity's sake). Indeed triangle areas in hyperbolic geometry are bounded and ideal triangles (with their three points sent to infinity) have the maximal area of $\pi \kappa^2$. This is very different from the flat case, which have of course unbounded triangle areas. 
\end{enumerate}
Let us detail those claims. The first statement reflects simply the fact that the construction of the three ``translation vectors'' $\ell_{1,2,3}$ corresponding to the three triangle edges are invariant under translations of the three vertices:
\be
\ell_{a,b,c}\in\SB(2,\C)\overset{L\in\SB(2,\C)}\longrightarrow L\ell_{a,b,c}\in\SB(2,\C)
\qquad\Longrightarrow\quad
\left|
\begin{array}{lcl}
\ell_{1}=\ell_{a}^{-1}\ell_{b}&\longrightarrow &\ell_{1}\\
\ell_{2}=\ell_{a}^{-1}\ell_{c}&\longrightarrow &\ell_{2}\\
\ell_{3}=\ell_{b}^{-1}\ell_{c}&\longrightarrow &\ell_{3}
\end{array}
\right.\,.
\ee
As a consequence, the $\SU(2)$ group elements $h_{1,2,3}$, defined from the Cartan decomposition of the $\ell_{1,2,3}$ are also invariant under  global translation of the triangle. Let us point out that the group elements $\ell_{1,2,3}$ and $h_{1,2,3}$ are not invariant under $\SU(2)$ rotations. Indeed considering the action of $k\in\SU(2)$ on $\ell_{a,b,c}$, we don't directly a triangular matrix, we have to correct the action by some $\SU(2)$ group elements on the right and we have to play around a little bit to get a non-trivial action of rotations:
$$
\ell_{a,b,c}\overset{k\in\SU(2)}\longrightarrow \ell_{a,b,c}^{(k)}=k\ell_{a,b,c}k_{a,b,c}^{-1},
\quad
\textrm{with}
\,\,
\left\{\begin{array}{l}
\ell_{a,b,c}^{(k)}\in\SB(2,\C),\\
k_{a,b,c}\in\SU(2),
\end{array}\right.
\quad\Longrightarrow\quad
\left|
\begin{array}{lcl}
\ell_{1}=\ell_{a}^{-1}\ell_{b}&\longrightarrow &\ell_{1}^{(k)}=\ell_{a}^{(k)-1}\ell_{b}^{(k)}=k_{a} \ell_{1}k_{b}^{-1}\\
\ell_{2}=\ell_{a}^{-1}\ell_{c}&\longrightarrow &\ell_{2}^{(k)}=k_{a} \ell_{2}k_{c}^{-1}\\
\ell_{3}=\ell_{b}^{-1}\ell_{c}&\longrightarrow &\ell_{3}^{(k)}=k_{b} \ell_{3}k_{c}^{-1}
\end{array}
\right.
$$
This can be converted into laws of transformations of the boosts and $\SU(2)$ holonomies under rotations:
$$
\left|
\begin{array}{lcl}
\ell_{1}=b_{1}h_{1}&\longrightarrow &\ell_{1}^{(k)}=k_{a} \ell_{1}k_{b}^{-1}=(k_{a}b_{1}k_{a}^{-1})\,(k_{a}h_{1}k_{b}^{-1})\\
\ell_{2}=b_{2}h_{2}&\longrightarrow &\ell_{2}^{(k)}=k_{a} \ell_{2}k_{c}^{-1}=(k_{a}b_{2}k_{a}^{-1})\,(k_{a}h_{2}k_{c}^{-1})\\
\ell_{3}=b_{3}h_{3}&\longrightarrow &\ell_{3}^{(k)}=k_{b} \ell_{3}k_{c}^{-1}=(k_{b}b_{3}k_{b}^{-1})\,(k_{b}h_{3}k_{c}^{-1})
\end{array}
\right.
$$
Finally, $H=h_{2}h_{3}^{-1}h_{1}$ gets mapped to $H^{(k)}=k_{a} H k_{a}^{-1}$ and is transformed by conjugation by the rotation $k_{a}$ (and not the original rotation $k$) rooted at its origin point.

\smallskip

As for the geometrical interpretation of the holonomy $H$, and the related second and third properties listed above,  geometry, directions and angles are more easily seen in the pure boost representation using the Cartan decomposition.
Let us consider the triangle rooted at the origin $M_{a}=\id$, taking $\ell_{a}=\id$ and thus $\ell_{b}= \ell_{1}$ and $\ell_{c}=\ell_{2}$ for simplicity's sake. The two other vertices are then $M_{b}=\ell_{b}\ell_{b}^{\dagger}=\ell_{1}\ell_{1}^{\dagger}=B_{1}^{2}$ and $M_{c}=\ell_{c}\ell_{c}^{\dagger}=\ell_{2}\ell_{2}^{\dagger}=B_{2}^{2}$, and its third edge boost is given by the closure relation $B_{3}=B_{1}^{-1}B_{2}H$ with $H\in\SU(2)$.  We parametrize these pure boosts by decomposing them onto the Pauli matrices, keeping the indices implicit:
$$
B
\,=\,
\cosh \frac{\eta}{2} + \sinh \frac{\eta}{2} \hat{u} \cdot \overrightarrow{\sigma}\,,
\qquad
\eta\ge0,
\quad\hu\in\cS_{2}\,.
$$
The boost parameters $\eta_{1,2}$ give the (hyperbolic) length of the two edges connected to the origin, while the boost directions $\hat{u}_{1,2}$ are the tangent vectors to the triangle edges at the origin vertex. Similarly $\eta_{3}$, defined through $\tr B_{3}^{2}=\tr b_{3}^{2}=\tr \ell_{3}\ell_{3}^{\dagger}$, gives the hyperbolic length of the third edge.

We also parametrize the $\SU(2)$ group element $H$, in terms of its rotation angle $\theta$ and axis $\hv$:
\be
H
\,=\,
\cos \frac{\theta}{2} + i \sin \frac{\theta}{2} \hat{v} \cdot \overrightarrow{\sigma}\,,
\qquad
\theta\in[0,2\pi],
\quad\hv\in\cS_{2}\,.
\ee
Let us look at its product with an arbitrary boost $B$. Computing the trace of their product gives:
\be
\tr HB = \cosh \frac{\eta}{2} \cos \frac{\theta}{2} + i \sinh \frac{\eta}{2} \sin \frac{\theta}{2} \hu \cdot \hv\,,
\ee
so we can extract the scalar product $\hu \cdot \hv$ from the imaginary part of this trace. Let us apply this to our three pure boosts $B_{1,2,3}$ related by the closure constraint relation 
$B_2^{-1} B_1 B_3 = H$. This gives us directly the equalities:
\be
\label{BBproduct}
\left|
\begin{array}{lcl}
B_2 H &=& B_1 B_3\\
 H B_3^{-1}&=& B_2^{-1} B_1\\
H^{-1}B_{1} &=& H^{-1}B_{2}H B_3^{-1}\\
\end{array}
\right.
\ee
Taking the trace of these matrix equalities and taking into account that pure boosts are Hermitian matrices and that the trace of two pure boosts is thus real, we have:
\be
\im\tr H^{-1}B_{1}
=
\im\tr B_2 H
=
\im\tr  H B_3^{-1}
=0\,,
\ee
thus implying that $\hv$ is orthogonal to all three boost directions $\hu_{1,2,3}$, or abusively that the rotation $H$ is normal to the three pure boosts $B_{1,2,3}$.
In particular, the rotation axis is collinear to the cross product $\hat{u}_{1}\times\hat{u}_{2}$ and the three boost directions $\hu_{1,2,3}$ are co-planar. This could be also checked by a full explicit calculation from $B_2^{-1} B_1 = H B_3^{-1}$,
computing explicitly $\eta_{3},\theta$ and $\hu_{3},\hv$ in terms of $\eta_{1,2}$ and $\hu_{1,2}$:
\begin{eqnarray}
\cosh \frac{\eta_1}{2} \cosh \frac{\eta_2}{2} - \sinh \frac{\eta_1}{2} \sinh \frac{\eta_2}{2}
 \hat{u}_1 \cdot \hat{u}_2
&=&
\cos \frac{\theta}{2} \cosh \frac{\eta_3}{2} \\
\sinh \frac{\eta_1}{2} \sinh \frac{\eta_2}{2} \hat{u}_1 \times \hat{u}_2 &=& \sin \frac{\theta}{2} \cosh \frac{\eta_3}{2} \hat{v}
\end{eqnarray}
Taking into account that the $\eta$'s are all positive, the first equation tells us that $\cos \frac{\theta}{2}$ is positive, thus  the angle $\theta$ actually lives in $[0,\pi]$ and the half-angle $\theta/2$ lives in $[0,\pi/2]$. This means that the normal rotation $H$ does run in the whole $\SU(2)$ group, but only in its $\SO(3)$ hemisphere connected to the identity. Further taking the ratio of the two equations, we have:
\begin{equation}
\label{rotH}
\left(
\frac{1}{\frac{1}{\tanh \frac{\eta_1}{2} \tanh \frac{\eta_2}{2}} - \hat{u}_1 \cdot \hat{u}_2}
\right)\,
\hat{u}_1 \times \hat{u}_2 = \tan \frac{\theta}{2} \hv\,,
\end{equation}
from which we can extract both angle and rotation axis. The pre-factor is always positive, since $1/\tanh$ is greater than 1 while the scalar product $\hat{u}_1 \cdot \hat{u}_2$ is always bounded by 1. The potential sign coming from the relative orientation of the two boost directions entering the cross product $ \hat{u}_1 \times \hat{u}_2$ is to be fully absorbed in the rotation axis $\hv$.

\medskip

Geometrically, at the origin $M=\id$, the tangent hyperplane $\cP^{0}$ to the hyperboloid $\cH_{3}$ is the 3d space orthogonal to the unit timelike vector $\cN^{0}=(1,0,0,0)$ and can be straightforwardly identified to $\R^{3}$. The isomorphism between $\cP^{0}$ and $\R^{3}$ is simply:
$$
\vx\in \R^{3} \leftrightarrow (0,\vx)\in\cP^{0}\subset\R^{3,1}\,.
$$
The tangent direction $\hat{u}_{1,2}$ live in this space. The 3d direction normal to the triangle is naturally the direction $\vN=\hat{u}_{1}\times\hat{u}_{2}$ orthogonal to both $\hat{u}_{1}$ and $\hat{u}_{2}$, and thus collinear to $\hv$. Moreover, we have:
$$
\tr (\vN\cdot\vsigma)(\cosh \frac{\eta}{2} + \sinh \frac{\eta}{2} \hat{u}_{a}\cdot\vsigma )=
\tr (\vN\cdot\vsigma)(\cosh \frac{\eta}{2} + \sinh \frac{\eta}{2} \hat{u}_{b} \cdot\vsigma)=
0,
\quad\forall \eta\,,
$$
implying that both hyperbolic edges are orthogonal to the 4-vector $(0,\vN)$. We can go further and show that the third edge, and thus the whole triangle, are also orthogonal to this vector $(0,\vN)$. Therefore, the triangle $T$ entirely lives in the 2-hyperboloid defined as the intersection $\cH_{2}^{\cN}=\cH_{3}\cap\cP_{\cN}$ of the 3-hyperboloid and the hyperplane $\cP_{\cN}$ orthogonal to $\cN=(0,\vN)$. The 2-hyperboloid $\cH_{2}^{\cN}$ is special, it contains the time-like vector $(1,0,0,0)$ and is a section of the 3-hyperboloid defined by merely removing one spatial direction. It is possible to check by a straightforward calculation\footnotemark that the points along the third edge of the triangle also lay in  $\cH_{2}^{\cN}$. Actually this is true whatever the lengths $\eta_{1,2}$ of the two edges attached to the origin, and all the hyperbolic triangles with one vertex at the origin and two vertices along the boost directions $\hu_{1}$ and $\hu_{2}$ all lay in this 2-hyperboloid $\cH_{2}^{\cN}$.
\footnotetext{The points of the third edge $3$ can defined by the composed boosts $\Lambda_{\eta}=B_{1}(\cosh\f\eta2+\sinh\f\eta2 \hu_{3}\cdot\vsigma)$ where $\hu_{3}$ is the boost direction of $B_{3}$ and the boost parameter $\eta$ runs from 0 to $\eta_{3}$. We can then check that these points all lay in the hyperplane $\cP_{\cN}$ by computing the scalar product $\tr (\vN\cdot\vsigma) \Lambda_{\eta}\Lambda_{\eta}^{\dagger}$.
}

\medskip

\begin{figure}[h!]
\includegraphics[angle=-90,scale=0.7]{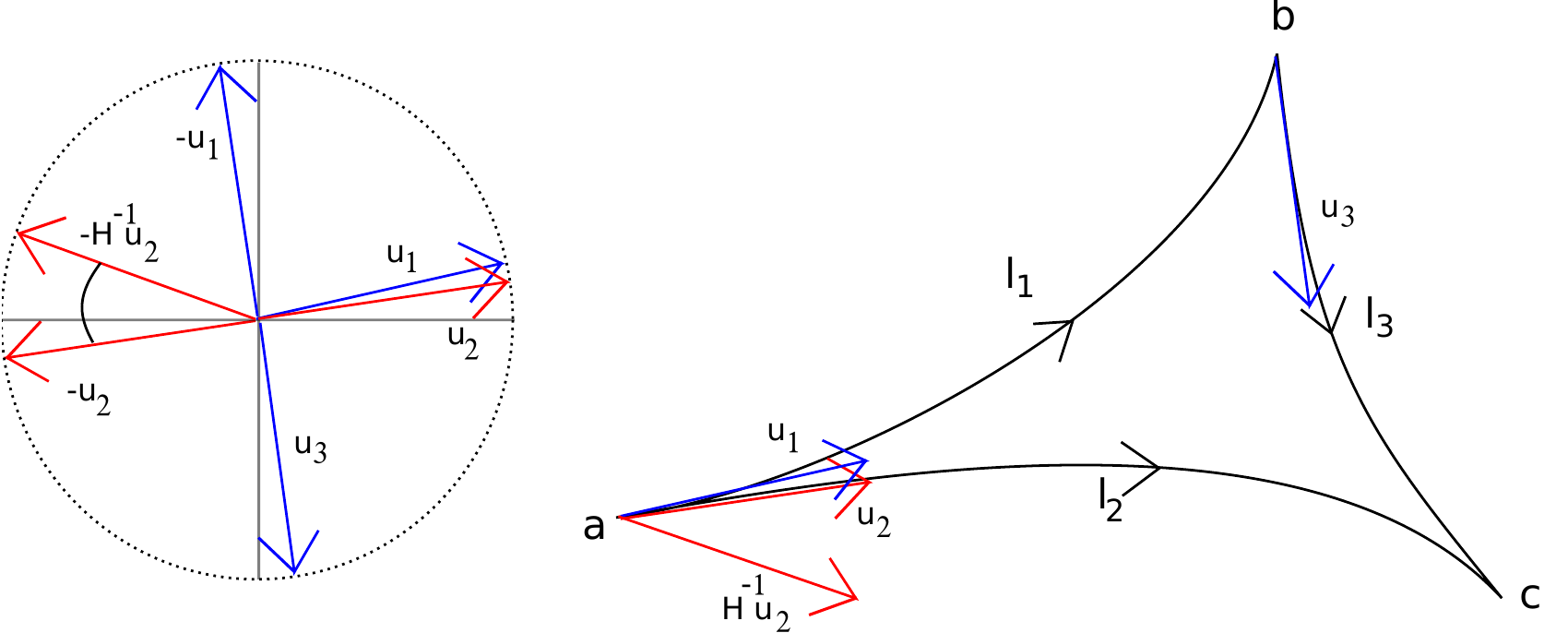}
\caption{Three vectors considered with the opposite of the first one always complete a $\pi$ angle. The rotation $H$ thus induces the deficit angle around the hyperbolic triangle.}
\label{fig:Deficit}
\end{figure}

Now that we have related the rotation axis $\hv \propto \hu_{1}\times \hu_{2}$ of the $\SU(2)$ holonomy $H$ to the normal direction to the triangle, we can further relate its rotation angle $\theta$ to the deficit angle around the hyperbolic triangle, and thus to its (hyperbolic) area (up to the $\ka^{2}$ factor that we have set to 1 for the sake of simplicity). To this purpose, we extract the cosine laws of the hyperbolic triangle from the closure relation $B_2^{-1} B_1 B_3 = H$. More precisely, we take the trace of the Hermitian square of the three boost products given above in \eqref{BBproduct}:
\begin{eqnarray}
\label{BBtrace}
\mathrm{Tr} B_2^2 & =& \mathrm{Tr} \left(B_1^2 B_3^{-2}\right)\nn \\
\mathrm{Tr} B_3^2 & =& \mathrm{Tr} \left(B_1^2 B_2^2\right)  \\
\mathrm{Tr} B_1^2 & =& \mathrm{Tr}\left( B_3^2 H^{-1} B_2^2 H \right)\,.\nn
\end{eqnarray}
And we compare the resulting equalities to the cosine law for a hyperbolic triangle:
\begin{eqnarray}
\cosh \eta_1 &=& \cosh \eta_2 \cosh \eta_3 - \sinh \eta_2 \sinh \eta_3 \cos \theta_{23} \nn\\
\cosh \eta_2 &=& \cosh \eta_1 \cosh \eta_3 - \sinh \eta_1 \sinh \eta_3 \cos \theta_{13} \\
\cosh \eta_3 &=& \cosh \eta_1 \cosh \eta_2 - \sinh \eta_1 \sinh \eta_2 \cos \theta_{12}\,,\nn
\end{eqnarray}
relating the hyperbolic edge length $\eta_{1,2,3}$ to the triangle angles $\theta_{c}=\theta_{23}$, $\theta_{b}=\theta_{13}$ and $\theta_{a}=\theta_{12}$ (between these edges).
By definition of the boosts  and triangle, we know that the edge lengths are given by the boost parameters $\cosh \eta_i = \mathrm{Tr} B_i^2$. Then the trace equalities allow to compute the triangle angles from the scalar products between the boost direction. The only non-trivial data is the action by conjugation by the $\SU(2)$ holonomy $H$ in the last identity above in \eqref{BBtrace}, which affect the definition of the angle $\theta_{bc}$ opposite to the edge $a$:
%
%
\begin{equation}
\cos \theta_{13} = - \hat{u}_1 \cdot \hat{u}_3,~
\cos \theta_{12} = \hat{u}_1 \cdot \hat{u}_2,~
\cos \theta_{23} = \left( H^{-1} \triangleright \hat{u}_2\right) \cdot \hat{u}_3\,,
\end{equation}
where $H$ acts as a 3d rotation on the boost direction $\hu_{2}$.
If $H$ was the identity $\id$, this would be proof that the angles sums up to $\pi$.
But here, for a non-degenerate hyperbolic triangle, the $\SU(2)$ holonomy $H$ is generically non-trivial and its rotation angle is thus exactly the deficit angle $\pi-(\theta_{23}+\theta_{13}+ \theta_{12})$ around the triangle. Since the hyperbolic triangle area is given by the deficit angle (up to the scale factor $\ka^{2}$), we conclude that the $\SU(2)$ group element $H$ encodes  the area of the triangle (fig.\ref{fig:Deficit}).

To conclude this section on the hyperbolic triangle, we have defined the translational invariant $\SU(2)$ holonomy $H$ around the triangle, by taking the product of the $\SU(2)$ group elements of the Cartan decomposition of the three translations defining the three edges of the triangle. This holonomy $H$ is to be interpreted as the oriented normal rotation to the triangle. Its rotation angle is bounded by $\pi$ and the deficit angle gives the area of the triangle, while the rotation axis defines the normal to the triangle at the origin point of the triangle and by extension the hyperplane section of the 3-hyperboloid in which it lives. 

In the next section, we will consider four triangles forming a hyperbolic tetrahedron and consider the closure relation satisfied by the four oriented (outward) normal rotations to these triangles.


\section{Compact closure constraint for the hyperbolic tetrahedron}

\subsection{Closure constraint for the normal rotations}

\begin{figure}[h!]
\includegraphics[scale=0.8,angle=-90]{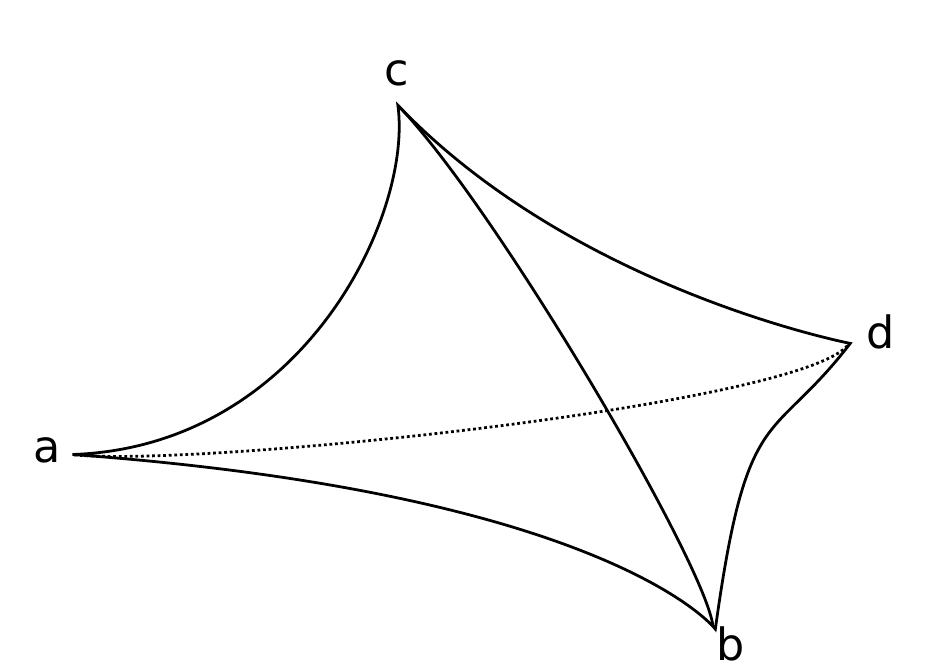}
\caption{We consider a tetrahedron defined by its four vertices.}
\label{fig:hypertetra}
\end{figure}

After having studied in great details the algebra and geometry of the hyperbolic triangle embedded in the 3-hyperboloid $\cH_{3}$, we  finally consider the tetrahedron in hyperbolic space. It is defined by four points, $a$, $b$, $c$ and $d$ (fig.\ref{fig:hypertetra}), defined by their respective triangular matrix $\ell_{a,b,c,d}$ describing their position on the hyperboloid.

We want to find a closure constraint for this hyperbolic tetrahedron in terms of the triangle normals that we  defined above. On such a tetrahedron, each edge carries a translation vector on the hyperboloid, defined as a $\SB(2,\C)$ group element, $\ell_{ab}=\ell_{a}^{-1}\ell_{b}$ and so on. We Cartan-decompose each translation to in order to identify their pure boost and rotation components, $\ell_{ab}=b_{ab}h_{ab}$. This requires choosing an orientation for all the edges. We fix the orientation of the tetrahedron by  alphabetic order, so the edge $ab$ is from $a$ to $b$ and so on. Switching the orientation of an edge simply amounts of inverting the translation, which leads to switching the rotation to its inverse while the boost gets braided:
$$
\ell_{ab}=\ell_{a}^{-1}\ell_{b}=b_{ab}h_{ab}
\qquad
\longrightarrow
\quad
\ell_{ba}=\ell_{b}^{-1}\ell_{a}=\ell_{ab}^{-1}
=(h_{ab}^{-1}b_{ab}^{-1}h_{ab})\,h_{ab}^{-1}\,.
$$
Here we will focus on the rotations and $\SU(2)$ holonomies, so it is enough to keep in mind that reversing the orientation simply inverts the $\SU(2)$ group elements.

\begin{figure}[h!]
{\includegraphics[scale=0.5,angle=-90]{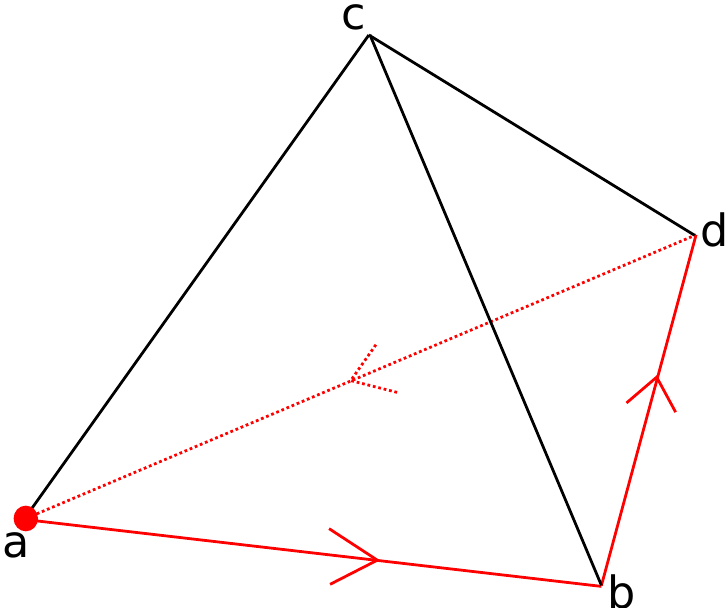}}
\qquad
{\includegraphics[scale=0.5,angle=-90]{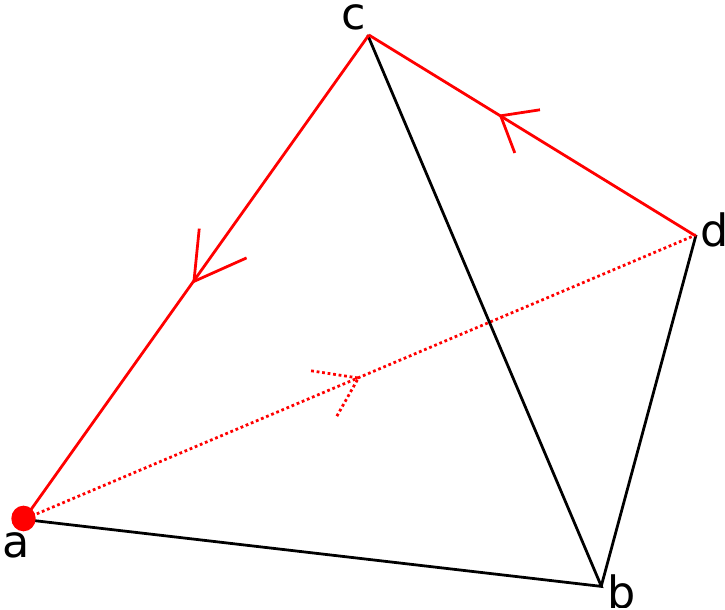}}
\qquad
{\includegraphics[scale=0.5,angle=-90]{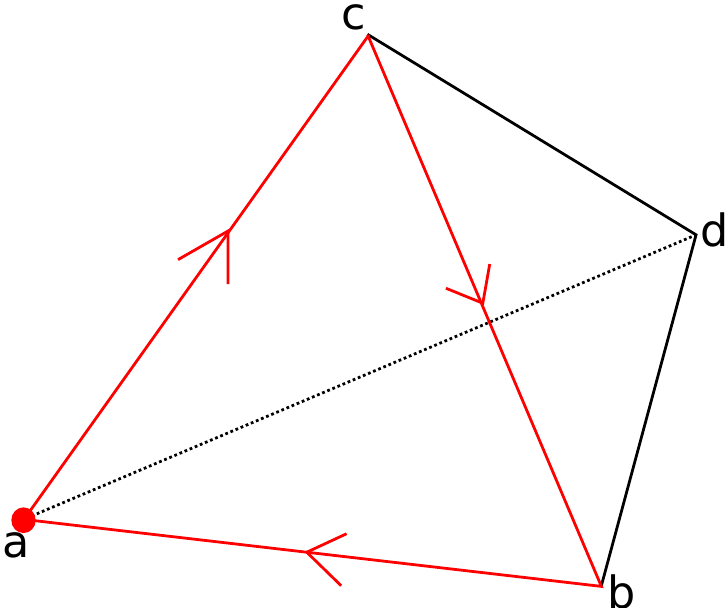}}
\qquad
{\includegraphics[scale=0.5,angle=-90]{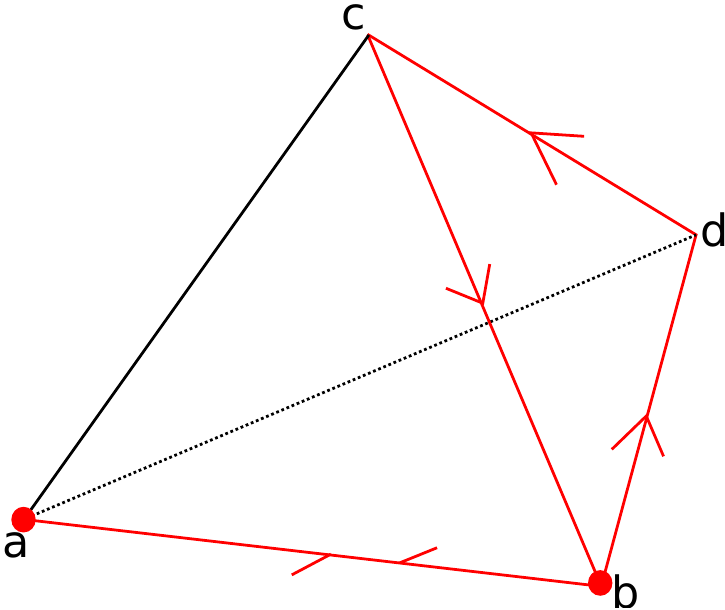}}

\caption{We define the $\SU(2)$ holonomies around the four triangles. We root all of them at the reference point $a$, which means that the holonomy of the fourth triangle must be parallel transported from point $b$ to point $a$. The $\SU(2)$ group elements along the edges compensate and the product of the four holonomies is constrained to the identity.}
\end{figure}

Let us choose a point on the tetrahedron, for instance $a$. This reference point will be taken as the origin point, or root, of the hyperbolic tetrahedron.
Starting from this point, we consider the three triangles sharing this node, $(abc)$, $(acd)$ and $(adb)$, and we  construct the three $\SU(2)$ holonomies defining the normal rotations to these hyperbolic triangles:
\begin{eqnarray}
H_{abc} &=& h_{ac} h_{bc}^{-1} h_{ab}^{-1}=h_{ac}h_{cb}h_{ba} \\
H_{acd} &=& h_{ad} h_{cd}^{-1} h_{ac}^{-1} \nn\\
H_{adb} &=& h_{ab} h_{db}^{-1} h_{ad}^{-1}=h_{ab} h_{bd} h_{ad}^{-1}\,,\nn
\end{eqnarray}
%
%
where we use the same convention for these three triangles as we have defined in the previous section on hyperbolic triangles, but for which we could easily switch orientation conventions without causing any issue.
We finally introduce the normal rotation to the fourth triangle $(bcd)$:
\be
H_{bcd} = h_{bd} h_{cd}^{-1} h_{bc}^{-1}\,.
\ee
It satisfies a simple closure relation:
\be
H_{adb} H_{acd} H_{abc} = h_{ab} H_{bcd} h_{ab}^{-1}\,,
\qquad
\boxed{H_{adb} H_{acd} H_{abc} \left(h_{ab} H_{bcd}^{-1} h_{ab}^{-1}\right) = \id}
\ee
The conjugation by the $\SU(2)$ group element $h_{ab}$ ensures that the holonomy around the fourth triangle is rooted at the same point $a$ as the other holonomy, in order to guarantee the same transformations for all under global Lorentz transformations.
From the point of view of differential geometry, the holonomies are the discrete counterparts of curvature and this closure relation is just Bianchi identity.

Finally, we have to settle the convention of ``inward'' vs ``outward'' normals to the triangles. Using the same conventions and definitions for the triangles  $(abc)$, $(acd)$ and $(adb)$, as in the previous section on hyperbolic triangles. The edge $(ab)$ is defined by the boost parameter $\eta_{ab}>0$ and boost direction $\hu_{ab}$, and similarly for the edges $(ac)$ and $(ad)$. Then the direction axis of the normal rotation $H_{abc}$ goes as $\hu_{ab}\times\hu_{ac}$. Everything about orientations happens exactly as in the flat case. We choose the order of the points $(a,b,c,d)$ such that $(\hu_{ab}\times\hu_{ac})\cdot\hu_{ad}\,<0$. This ensures that $H_{abc}$, $H_{acd}$ and $H_{adb}$ do define the outward normals to the there hypberbolic triangle sharing the root point $a$. Finally, $H_{bcd}$'s rotation axis goes along  $\hu_{bc}\times\hu_{bd}$, so that $H_{bcd}^{-1}$ defines the outward normal to the fourth triangle $(bcd)$. 

We can therefore conclude that, assuming the orientation condition $(\hu_{ab}\times\hu_{ac})\cdot\hu_{ad}\,<0$, our closure relation defines an identity satisfied by the four outward normal rotations to the hyperbolic triangles forming the tetrahedron. If this condition is not satisfied, we can simply swap the role of the points $b$ and $c$ and repeat the same procedure.


\subsection{Reconstructing the Hyperbolic Tetrahedron}

We want this closure condition on the normal rotations to uniquely define (up to a translation) a tetrahedron. Or in other words, we need to be able to reconstruct the original tetrahedron from the four normal rotations satisfying the closure relation.

Let us start with four $\SU(2)$ group elements $H_{A,B,C,D}$ satisfying the closure relation:
$$
H_{C}H_{B}H_{D}H_{A}=\id\,.
$$
We would like to identify these four group elements to the four normal rotations to the triangles of a hyperbolic tetrahedron. We first insist on two important remarks:
\begin{itemize}

\item We break the apparent symmetry between these four group elements and we single out the last group element $H_{A}$. We will identify the first three group elements $H_{B,C,D}$ to the normal rotations of to the three triangles sharing the root point of the tetrahedron. The last group element $H_{A}$ will be identified a posteriori as the holonomy around the last triangle (opposite to the root) properly pulled back to the root point. Our reconstruction procedure is thus not invariant under circular permutations of the four normal rotations. It might be interesting to study the effect of permutations (and the associated braiding) on the reconstructed tetrahedron, but it does not seem important for our present purpose and we postpone it for possible future investigation.

\item The four group elements  $H_{A,B,C,D}$ can not be arbitrary $\SU(2)$ elements. They are restricted to live in the $\SO(3)$ hemisphere connected to the identity, or in simpler terms $\tr H \ge 0$. This ensures that their rotation angles $\theta$ remains in $[0,\pi]$. We speculate that the other part of $\SU(2)$ will be relevant to spherical triangles and tetrahedra (embebbed in $\cS_{3}$).

\end{itemize}

\begin{figure}[h!]
\includegraphics[scale=0.85,angle=-90]{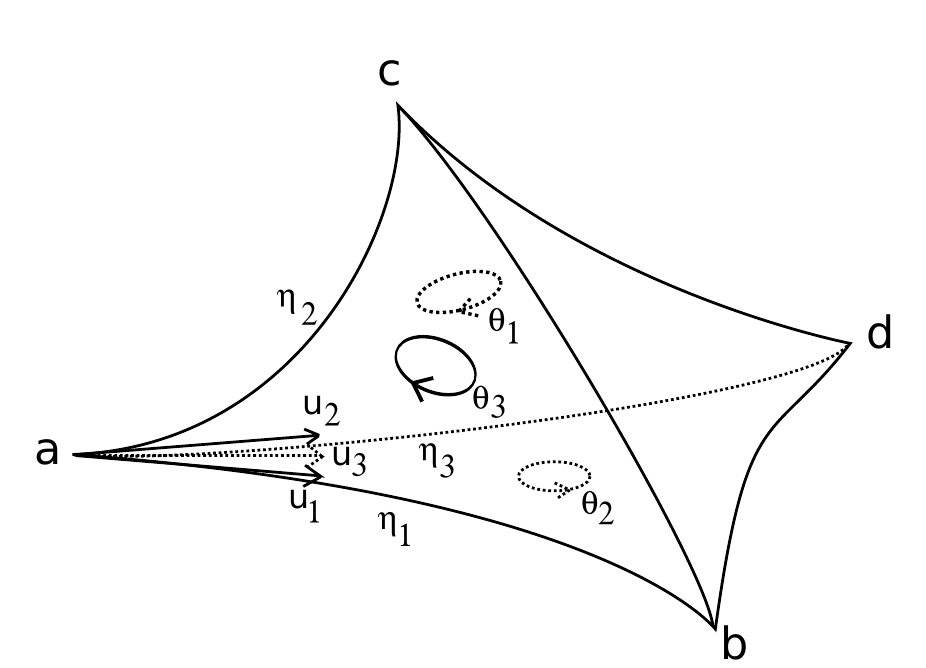}
\caption{Notations for the reconstruction of the hyperbolic tetrahedron}
\label{fig:recons}
\end{figure}

We now proceed similarly to the last case. We first fix the root $a$, or apex, of our tetrahedron to the hyperboloid origin for simplicity's sake. From the three normal rotations to the triangles attached to $a$, we can deduce the boost directions $\hat{u}_1$, $\hat{u}_2$ and $\hat{u}_3$ of the three edges leaving $a$ towards the points $b,c,d$. They are the tangent vectors of these edges. We simply have to determine the lengths of those edges to conclude the reconstruction. In the flat case, we used the tetrahedron volume as a global scale factor to compute the edge lengths. Here we will proceed slightly differently and use the deficit angles associated to the three triangles, $\theta_1$, $\theta_2$ and $\theta_3$, (see fig.\ref{fig:recons}), to deduce the edge lengths, $\eta_1$, $\eta_2$ and $\eta_3$.

%
%
%

For the triangle $(abc)$, the holonomy angle $\theta_{3}$ and the normal rotation axis $\hv_{3}$ are related to the boost parameters and directions through eqn.\eqref{rotH}:
$$
\left(
\frac{1}{\frac{1}{\tanh \frac{\eta_1}{2} \tanh \frac{\eta_2}{2}} - \hat{u}_1 \cdot \hat{u}_2}
\right)\,
\hat{u}_1 \times \hat{u}_2 = \tan \frac{\theta_{3}}{2}\, \hv_{3}\,,
$$
Since all the factors are positive by definition or construction, one can get the direction $\hu_{1}$ by taking the cross-product $\hv_{3}\times \hv_{2}$:
$$
(\hv_{3}\times \hv_{2}) \propto ( \hat{u}_1 \cdot \hat{u}_2) \times ( \hat{u}_3 \cdot \hat{u}_1)
\propto [\hat{u}_1\cdot (\hat{u}_3\cdot \hat{u}_2)]\,\hat{u}_1\,,
$$
keeping in mind that we have chosen the convention $\hat{u}_1\cdot (\hat{u}_3\cdot \hat{u}_2)<0$. to ensure that we are looking at the outward triangle normals.
Now that we have reconstructed the direction $\hu_{1,2,3}$ from the normal rotations, we are left with computing the edge lengths $\eta_{1,2,3}$.

Let us call $\theta_{12}\in[0,\pi]$ the angle between the boost directions $\hat{u}_1$ and $\hat{u}_2$, with the convention $\sin\theta_{12}\ge 0$. Then we can express the product of the hyperbolic tangents of $\eta_{1}$ and $\eta_{2}$ in terms of $\theta_{3}$ and $\theta_{12}$. Proceeding similarly for the two other triangles gives us the following identities:
\begin{eqnarray}
\tanh \frac{\eta_1}{2} \tanh \frac{\eta_2}{2} &=& \frac{\tan \frac{\theta_3}{2}}{\sin \theta_{12} + \tan \frac{\theta_3}{2} \cos \theta_{12}} \\
\tanh \frac{\eta_1}{2} \tanh \frac{\eta_3}{2} &=& \frac{\tan \frac{\theta_2}{2}}{\sin \theta_{13} + \tan \frac{\theta_2}{2} \cos \theta_{13}} \\
\tanh \frac{\eta_2}{2} \tanh \frac{\eta_3}{2} &=& \frac{\tan \frac{\theta_1}{2}}{\sin \theta_{23} + \tan \frac{\theta_1}{2} \cos \theta_{23}}
\end{eqnarray}
%
Taking the appropriate products and ratios, we get:
\begin{eqnarray}
\tanh^2 \frac{\eta_1}{2} &=& \frac{\sin (\theta_{23} - \frac{\theta_1}{2})}{\sin \frac{\theta_1}{2}} \times \frac{\sin \frac{\theta_2}{2}} {\sin (\theta_{13} - \frac{\theta_2}{2})} \times \frac{\sin \frac{\theta_3}{2}} {\sin (\theta_{12} - \frac{\theta_3}{2})} \\
\tanh^2 \frac{\eta_2}{2} &=& \frac{\sin (\theta_{13} - \frac{\theta_2}{2})}{\sin \frac{\theta_2}{2}} \times \frac{\sin \frac{\theta_1}{2}} {\sin (\theta_{23} - \frac{\theta_1}{2})} \times \frac{\sin \frac{\theta_3}{2}} {\sin (\theta_{12} - \frac{\theta_3}{2})} \\
\tanh^2 \frac{\eta_3}{2} &=& \frac{\sin (\theta_{12} - \frac{\theta_3}{2})}{\sin \frac{\theta_3}{2}} \times \frac{\sin \frac{\theta_2}{2}} {\sin (\theta_{13} - \frac{\theta_2}{2})} \times \frac{\sin \frac{\theta_1}{2}} {\sin (\theta_{23} - \frac{\theta_1}{2})}
\end{eqnarray}
Thus positive solutions are unique if they exist. So we proved the unicity of the solution. We are however left with a priori non-trivial constraints between the initial normal rotations in $\SU(2)$. At this stage, we do not have a clear geometrical understanding of these constraints, although they are clearly related to triangular inequalities between the angles within each triangle. When one of these signs is negative, it is equivalent of an obtuse angle in one of the triangle, which should be compensated by at least another angle becoming obtuse too. We could conjecture that these sign constraints simply amount to the initial constraint that $\tr H_{A}>0$, i.e that the fourth triangle reconstructed a posteriori is also a well-defined hyperbolic triangle. However, it could also lead to an additional constraint, for instance on the Gram matrix between the holonomies as suggested in \cite{Haggard:2014xoa}.

We expect the sign violating sector to correspond to spherical triangles and tetrahedron, in which case the same  type of $\SU(2)$ closure constraints would allow to represent tetrahedra in both positive and negative curvature. We understand that work in this direction has been recently presented in \cite{Haggard:2014xoa}.

\section{Non-compact closure constraint for the hyperbolic tetrahedron}

\subsection{From a compact to non-compact closure constraint}

As for now, we have identified a closure relation for the hyperbolic tetrahedron in terms of normals, associated each of its triangular face, living in the compact group $\SU(2)$. The geometrical interpretation of this closure relation is clear but it is  algebraically very differently from the Gauss law proposed in \cite{HyperbolicPhaseSpace}, living on graph's vertices generating the $\SU(2)$ gauge invariance, which should be related to a closure constraint for hyperbolic polyhedra in the expected curved twisted geometry interpretation of the deformed phase space for loop quantum gravity. Indeed, the Gauss law reads as a product of non-compact $\SB(2,\C)$ group elements, $L_{A}L_{B}L_{C}L_{D}=\id$ for instance in our 4-valent case dual to a tetrahedron. These triangular matrices are understood as the hyperbolic counterpart of 3-vectors in the flat case, and thus are natural candidates to be understood as normal vectors to the triangular faces of the hyperbolic tetrahedron. However, up to now, we have develop a well-defined notion of ``normal rotations'' in $\SU(2)$ to the hyperbolic triangles, which seems natural taking into account that their area is given by the deficit angle and is thus bounded by $\pi$ (times the curvature scale $\ka^2$), and therefore naturally represented by a compact group element. The goal of this section is to find a common ground to these two points of view and identify a closure relation for the hyperbolic tetrahedron in terms of $\SB(2,\C)$ group elements and which would be a fortiori equivalent to the compact closure relation that we have already introduced and discussed in the previous section.

\smallskip

A first attempt towards this reconciliation is to try to directly convert the $H$-closure relation into triangular matrices. Indeed, considering the $\SU(2)$ holonomy $H$ around a hyperbolic triangle, it is constructed as a product $H = h_1 h_2 h_3$ of $\SU(2)$ group elements $h_{1,2,3}$ coming from the Cartan decomposition of the translation vectors $\ell_{1,2,3}$ along the three triangle edges. It is possible to write the  $\SU(2)$ group elements explicitly in terms of the original $\SB(2,\C)$ elements:
\begin{equation}
h_i = \frac{\ell_i + (\ell_i^\dagger)^{-1}}{\sqrt{2 + \mathrm{Tr} (\ell_i \ell_i^\dagger)}}\,.
\end{equation}
This allows to write the $\SU(2)$ normal rotation to the triangle directly in terms of the $\ell_{i}$'s:
\begin{equation}
H = \frac{1}{\sqrt{\left(2 + \mathrm{Tr} \ell_1 \ell_1^\dagger\right)\left(2 + \mathrm{Tr} \ell_2 \ell_2^\dagger\right)\left(2 + \mathrm{Tr} \ell_3 \ell_3^\dagger\right)}}\left[\left(2 + \mathrm{Tr} \ell_1 \ell_1^\dagger + \mathrm{Tr} \ell_3 \ell_3^\dagger \right) + (\ell_1 \ell_1^\dagger) (\ell_3^{-1} (\ell_3^{-1})^\dagger)^{-1} + (\ell_1 \ell_1^\dagger)^{-1} (\ell_3^{-1} (\ell_3^{-1})^\dagger) \right]
\end{equation}
which was simplified using the triangle closure relation $\ell_1 \ell_2 \ell_3 = 1$.
Since we can express the normal rotations in terms of the $\SB(2,\C)$ group elements, we can translate the whole compact closure relation for the tetrahedron in terms of triangular matrices. However, without some magic trick or simplification, the resulting relation does not look like a simple equation $L_{A}L_{B}L_{C}L_{D}=\id$ in terms of $\SB(2,\C)$ normal vectors to the triangles.

\smallskip

It might be more enlightening to look at these closure relations from a geometrical point of view. Our hyperbolic tetrahedron is defined by 4 points on the 3-hyperboloid $\cH_{3}$. This also defines a flat tetrahedron in the four-dimensional $\R^{3,1}$. The hyperbolic tetrahedron has bounded hyperbolic triangle areas while the flat tetrahedron has naturally unbounded flat triangle areas. Let us look at the flat tetrahedron, each face $i$ has a normal bi-vector $B^{\mu\nu}_i$ (antisymmetric in $\mu\leftrightarrow\nu$) satisfying the bi-vector closure constraints:
\be
\sum_i B^{\mu\nu}_i = 0\,.
\ee
To turn it into a 3-vector closure constraint, of the type $\sum_{i} \vN_{i}=0$ with $\vN_{i}\in\R^3$, we can identify the (time-like) vector $\cT$ normal  to the hyperplane containing the flat tetrahedron. But then to reconstruct the tetrahedron in $\R^{3,1}$, we would require both the 3-vector closure constraint and the time-like normal 4-vector $\cT$. Another way to proceed is to focus on the spatial part of the bivectors $B^{ab}_i$, that is their projection on the canonical $\R^3$ hyperplane. A 3-bivector carries the same data as a 3-vector and we are left with a spatial closure constraint  $\sum_{i} {B}^c_{i}=0$, which defines a unique tetrahedron in $\R^3$. Now to recover the original tetrahedron in the four-dimensional space-time $\R^{3,1}$, we simply have to lift each point from $\R^3$ to the hyperboloid $\cH_{3}$ by assigning them the appropriate (unique) time coordinate.This way, both the flat tetrahedron or hyperbolic tetrahedron are entirely determined by their (same) spatial projection, which defines a tetrahedron in $\R^3$.

However, this way of describing a hyperbolic tetrahedron by its 3d shadow defined by a classical flat 3-vector closure constraint is not the way we proceed here, although it might be an interesting point of view to bridge between the geometrical interpretation of regular twisted geometries and the want-to-be hyperbolic twisted geometries encoded by the deformed loop gravity phase space introduced in \cite{HyperbolicPhaseSpace}. Here, it seems that the appropriate perspective is that we do not consider the one hyperplane containing the whole flat tetrahedron, but the individual hyperplanes each containing one hyperbolic triangular face of the hyperbolic tetrahedron. Then we would be considering the 3d normal vector to each face but these different hyperplanes, which must be related to each other by a suitable Lorentz transformations. This would definitely lead to a deformed closure constraint for the hyperbolic tetrahedron, with some non-abelian structure (due to the non-trivial parallel transport between the reference frames of each hyperbolic triangle).

Although we keep this as intuition, we will not proceed along this path here. We choose a simpler method, directly defining $\SB(2,\C)$ normals to each hyperbolic triangle, such that they obviously satisfy a simple closure relation. Working at a completely algebraic level allows us to further show that our closure constraints reversely uniquely defines the hyperbolic tetrahedron (up to a global translation).

%
%
\begin{figure}[h!]
\includegraphics[angle=-90,scale=1.]{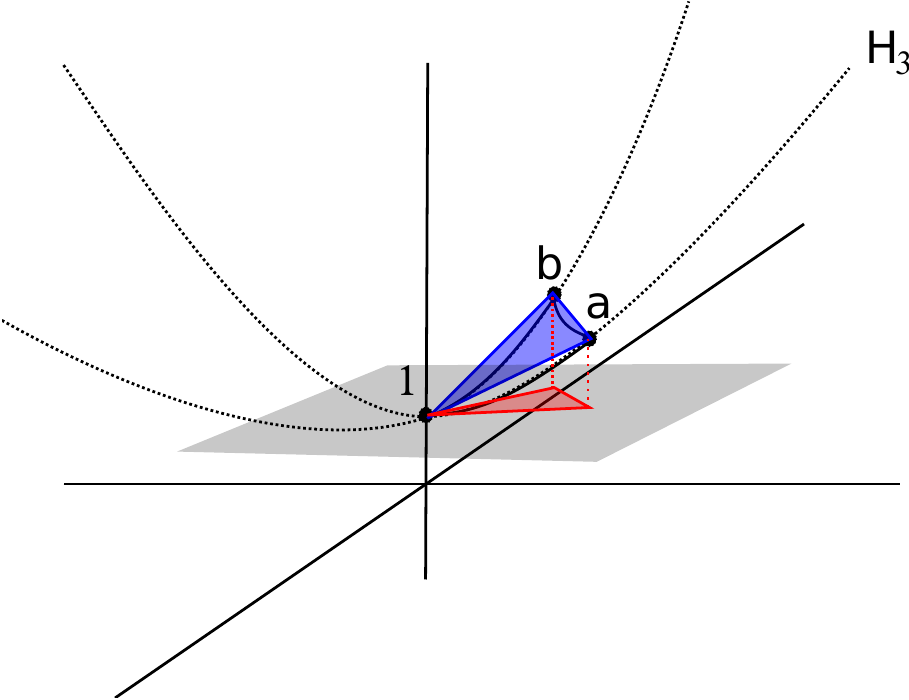}
\caption{Schematic view of the possible projections in 3D. In blue, we have the projection onto the hyperplane containing the flat tetrahedron, in green, the projection onto the tangent hyperplane to the identity.}
\end{figure}

\subsection{Non-compact closure constraint and reconstruction}

We consider a hyperbolic tetrahedron formed by the 4 points $a,b,c,d$, with the four triangles $(abc)$, $(bcd)$, $(cda)$ and $(dab)$.
Looking more closely at one triangle, say $(abc)$, its three vertices and defined by three translations $\ell_{a,b,c}$, from which we defined the hyperbolic translation vectors along the three edges as $\SB(2,\C)$ elements, which satisfy the triangle closure constraint:
\be
\left|
\begin{array}{l}
\ell_{ab}=\ell_{a}^{-1}\ell_{b}\\
\ell_{ac}=\ell_{a}^{-1}\ell_{c}\\
\ell_{bc}=\ell_{b}^{-1}\ell_{c}
\end{array}
\right.
\qquad
\textrm{satisfying}
\quad
\ell_{ab}\ell_{bc}=\ell_{ac}\,,
\quad
\textrm{or equivalently}
\quad
\ell_{ab}\ell_{bc}\ell_{ac}^{-1}=\id\,.
\ee
We define the $\SB(2,\C)$ normal to the triangle as:
\be
L_{abc}
\,\equiv\,
\ell_{ab}^{-1} \ell_{bc} 
\,=\,
\ell_{b}^{-1}\ell_{a}\ell_{b}^{-1}\ell_{c}\,.
\ee
Since the edge vectors $\ell_{ab}$ and $\ell_{bc}$ describe the edge displacements on the 3-hyperboloid, $L_{abc}$ is algebraically similar to a wedge product between edge vectors. Moreover, we could translate the $\ell$'s in terms of boosts, with boost parameters giving the edge lengths and boost directions giving the edge tangent directions, and this would allow to express $L_{abc}$ in terms of geometrical observables. But we haven't identified any simple geometrical interpretation similarly to the fact that the rotation angle of the normal rotation giving the area of the hyperbolic triangle. On the other hand, this definition of the $\SB(2,\C)$ normal to the triangle allows for a very simple non-compact closure constraint.

\begin{figure}[h!]
\includegraphics[scale=0.7,angle=-90]{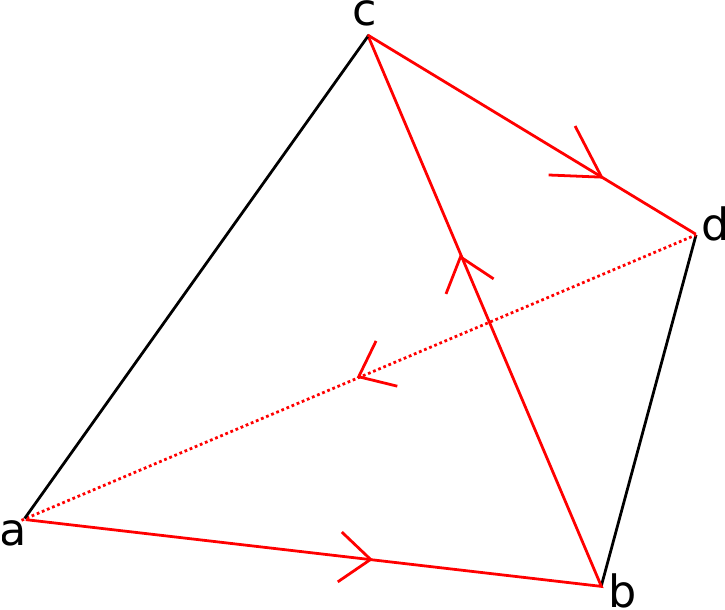}
\caption{The closure requires the choice of a path on the tetrahedron.}
\label{fig:path}
\end{figure}

Indeed, let us choose a definite path along the tetrahedron vertices (see fig.\ref{fig:path})., say $(abcd)$ following alphabetical order for the sake of simplicity, and let us consider the $\SB(2,\C)$ normal to the triangles following the path's order:
\begin{equation}
\left|\begin{matrix}
L_{abc} &=& \ell_{ab}^{-1} \ell_{bc} \\
L_{bcd} &=& \ell_{bc}^{-1} \ell_{cd} \\
L_{cda} &=& \ell_{cd}^{-1} \ell_{ad}^{-1} \\
L_{dab} &=& \ell_{ad} \ell_{ab}
\end{matrix}\right. ~~\quad \Large{\in SB(2,\C)}\normalsize{~}
\end{equation}
And these $\SB(2,\C)$ normals obviously satisfy the following closure relation by construction:
\begin{equation}
\boxed{L_{abc} L_{bcd} L_{cda} L_{dab} = \id.}
\end{equation}

\smallskip

Let us now prove that reversely this non-compact  closure relation in terms of $\SB(2,\C)$ group element uniquely determines the tetrahedron (up to global translations).
So let us start with four  $\SB(2,\C)$ group elements, $L_{abc}$, $L_{bcd}$, $L_{cda}$ and $L_{dab}$,  such as the closure condition is satisfied. We would like to reconstruct all the six edge translations, $\ell_{ab}$ and so on, from the $L$'s. Keeping in mind both the definition of the $\SB(2,\C)$ normals $L$'s and the triangle closure constraints satisfied by the $\ell$'s by definition, we can directly  reconstruct all the $\ell$'s from the $L$'s plus a single edge vector, say, $\ell_{ab}$. Indeed the solution must be of the following type, in terms of a single $\ell\in\SB(2,\C)$ left to determine:
\begin{equation}
\left\{\begin{matrix}
\ell_{ab} &=& \ell \\
\ell_{bc} &=& \ell L_{abc} \\
\ell_{cd} &=& \ell L_{abc} L_{bcd} \\
\ell_{ad} &=& L_{dab} \ell^{-1} \\
\ell_{ac} &=& \ell_{ab} \ell_{bc}=\ell^2 L_{abc} \\
\ell_{bd} &=& \ell_{bc} \ell_{cd}=\ell L_{abc} \ell L_{abc} L_{bcd} 
\end{matrix}\right.
\end{equation}
where $\ell_{bc}$ is given by the definition of the normal $L_{abc}$, $\ell_{cd}$ is then given by the definition of $L_{bcd}$, $\ell_{ad}$ is in turn deduced from the definition of $L_{dab}$, $\ell_{ac}$ is determined by the triangle closure constraint on $(abc)$ and finally $\ell_{bd}$ is obtained using the closure constraint on $(bcd)$.

There remain a single equation left to impose: the triangle closure constraint on $(cda)$ (or equivalently on $(dab)$). This translates to a non-linear equation on $\ell$:
\be
\ell_{cd}=\ell_{ca}\ell_{ad}=\ell_{ac}^{-1}\ell_{ad}
\qquad\Longrightarrow\quad
\ell L_{abc} L_{bcd} = L_{abc}^{-1} \ell^{-2}\,L_{dab} \ell^{-1}\,.
\ee
This equation can be written in a smoother way as:
\begin{equation}
\ell^{-1} L_{dab} \ell^{-1} L_{bcd}^{-1} = \ell L_{abc} \ell L_{abc}
\end{equation}
Writing this explicitly as 2$\times$2 triangular matrices, we find that there is a unique solution for $\ell$. The interested reader can see the derivation in details in the appendix \ref{app:derivation}. Explicitly, we get:
\begin{eqnarray}
\lambda &=& \left(\frac{\mu_{dab} \mu_{bcd}^{-1}}{\mu_{abc}^2}\right)^{\frac{1}{4}} \\
z &=& \frac{\lambda^2 \alpha_{bcd} - \mu_{bcd}^{-1}\alpha_{bad} - \mu_{abc} \alpha_{abc} - \lambda^{-2} \alpha_{abc} }{\lambda \mu_{abc}^2 + \lambda^{-1} (1 + \mu_{dab}\mu_{bcd}^{-1}) + \lambda \mu_{dab}^{-1} \mu_{bcd}^{-1}}
\end{eqnarray}
where we have parametrized the $\SB(2,\C)$ normals as
\begin{equation}
N_i = \begin{pmatrix}
\mu_i & 0 \\
\alpha_i & \mu_i^{-1}
\end{pmatrix}\,.
\end{equation}
Finally, one can check that this solution for $\ell$ does lead to a correct solutions for all the six edge translations, $\ell_{ab}$ and so on, satisfying the triangle closure constraints and defining the correct given $\SB(2,\C)$ normals.

This allows to conclude that our proposed non-compact closure condition does actually totally define the hyperbolic tetrahedron uniquely up to a global translation. It must be somehow equivalent to the compact closure constraints defined in terms of the normal rotation, although a direct link is not yet clear to us.


\section*{Conclusion \& Outlook}

In this paper, we investigated the question of closure constraints for the hyperbolic tetrahedron in the context of loop quantum gravity with a non-vanishing cosmological constant. Our goal was to do this first step towards interpreting the deformed phase space structure for loop gravity on a given graph defined in \cite{HyperbolicPhaseSpace}, by lifting the standard $T^*\SU(2)$ structure to $\SL(2,\C)$ provided with its classical $r$-matrix, as discrete 3d hyperbolic geometries to be embedded in a $3+1$-dimensional space-time.

The setting of our analysis is rather simple: we consider the hyperbolic tetrahedron formed by four points $(a,b,c,d)$ on the 3-hyperboloid of fixed norm time-like vectors in $\R^{3,1}$ with its four faces as hyperbolic triangles, and we look for a closure constraint between ``normals'' associated to its triangular faces that carry the information about their area and direction (for instance the 2-hyperboloid section or 3d hyperplane in which the triangle lives). For a flat tetrahedron in $\R^3$, the normal vectors to the triangle are 3-vectors, constrained to sum to 0. When providing the set of flat tetrahedra with an appropriate Poisson structure, this simple closure constraint is further understood to generate global 3d rotations. The task was to generalize this to the hyperbolic tetrahedron.

The simplest closure constraint for the hyperbolic tetrahedron turned out actually to consider the standard flat closure constraint for the flat tetrahedron made of the four projections of the hyperbolic tetrahedron's vertices onto the canonical hyperplane $\R^3$. This somewhat artificial method does not allow for an intrinsic understanding of the discrete hyperbolic geometry. Following the structures of the deformed phase space proposed in  in \cite{HyperbolicPhaseSpace}, we have been led to two more interesting proposals:
\begin{itemize}
\item A compact closure constraint, in terms of $\SU(2)$ group elements interpreted as the normal rotations to the hyperbolic triangles:
%
\begin{equation*}
H_{abc} H_{acd} H_{adb} \tH_{bdc}= \id\,,
\qquad
\tH_{bdc}=(h_{ab} H_{bdc} h_{ab}^{-1})=(h_{ab} H_{bcd} h_{ab}^{-1})^{-1} \,,
\end{equation*}
where the $H$'s are $\SU(2)$ holonomies around the triangular faces, defining the ``normal rotations'' to these hyperbolic triangles such that their rotation angle gives the triangle's area and their rotation axis is the normal to the plane tangent to the triangle at its chosen root point. The first three holonomies correspond to the three triangles sharing the root point $a$ of the tetrahedron, while the fourth holonomy needs to be corrected by a  conjugation by a group element $h_{ab}$ representing the parallel transport along the edge $ab$. There is another constraint on the normal rotations, $\tr H\ge 0$, which correspond to the fact that the sum of the angles around a hyperbolic triangle is bounded by $\pi$. It could further require an additional constraint, potentially on the Gram matrix of the holonomies $H$, to ensure the existence of the hyperbolic tetrahedron.
\item A non-compact closure constraint, in terms of $\SB(2,\C)$ group elements interpreted as the normal vectors to the hyperbolic triangles:
\begin{equation*}
L_{abc} L_{bcd} L_{cda} L_{dab} = \id\,,
\end{equation*}
where the $L$'s are $\SB(2,\C)$ elements similar to the wedge product of edge translation vectors for each hyperbolic triangle, mimicking the construction for the flat triangles and tetrahedron. This closure constraint depends on the choice of a (cyclic) order between the four vertices of the tetrahedron, which determines the order of the triangles appearing in the constraint and definition of each $L$ for each triangle.
\end{itemize} 

Both closure constraints uniquely determines the hyperbolic tetrahedron up to global translations on the 3-hyperboloid. Their offer their own perspective and have their own different advantages. On the one hand, the compact closure constraint has a neat geometrical interpretation and seems to have one sector corresponding to the hyperbolic tetrahedron and triangles, while another sector (with $\tr H\le 0$) would correspond to spherical tetrahedron and triangles, thus allowing for the possibility of switch of signs for the cosmological constant or more generally for a local average mean curvature. A similar point of view has been put forward and explored in the recent work \cite{Haggard:2014xoa}. On the other hand, the non-compact closure constraint has a incredibly simple algebraic structure and corresponds to the same Gauss law generating the $\SU(2)$ gauge invariance (equivalently 3d global rotations of the tetrahedron and triangles) in the deformed phase space  for loop quantum gravity as proposed in \cite{HyperbolicPhaseSpace}.

The present analysis could be further clarified by some mathematical improvements, such as a proof (or dis-proof) that there is indeed no further implicit constraint between the normal rotations to the hyperbolic triangles, or a clear geometrical interpretation for the $\SB(2,\C)$ normal vectors to the hyperbolic triangles, and finally a direct proof of the equivalence between the compact and non-compact closure constraints.
Beyond these technical issues, this work opens the door to deeper questions about the algebra and geometry of the $q$-deformed phase space for loop quantum gravity and the interpretation of the deformation as accounting for a non-vanishing cosmological constant:
\begin{itemize}

\item In the deformed phase space on a given graph $\Gamma$, one defines two types of constraints: a Gauss law at the graph's nodes, as a product of $\SB(2,\C)$ group elements, which generates a gauge invariance under $\SU(2)$ transformations, and a deformed flatness constraints around the graph's loops, as a product of $\SU(2)$ group elements, which generates translations of the corresponding elements. Following the standard interpretation of the loop gravity phase space as twisted geometries, it is therefore tempting and natural to seek to interpret the Gauss law at a closure constraints for discrete blocks of hyperbolic geometry, in particular for hyperbolic tetrahedron dual to 4-valent nodes.This is the perspective taken in the present work. One should nevertheless wonder if the geometrical interpretation of the Gauss law and flatness constraint is as straightforward as proposed in \cite{HyperbolicPhaseSpace} and whether one can not be mapped onto the other, or at least reformulated in similar terms.

\item Following this line of thought, we would like to recast the set of hyperbolic tetrahedra as a phase space, as in the flat case, and analyze the flow generated by the closure constraints introduced here. We expect the non-compact closure constraints to generate 3d rotations, while we do not expect the compact  closure constraints to be first class. Failure of this expectation might indicate a wrong intuition in the geometrical interpretation of the deformed phase space as hyperbolic twisted geometries. One should also keep in mind that, although it is convenient to interpret the loop gravity phase space on a fixed graph, and therefore its spin network states at the quantum level, as discrete classical geometries (twisted geometries, generalizing Regge triangulations), this is not necessarily an imperative of the theory.

\item Since the compact closure constraint seems to offer two sectors, one corresponding to hyperbolic tetrahedra, while the other is speculated to correspond to spherical tetrahedron, it would be interesting to investigate this direction further and understand if it is possible to define a phase space containing both spherical and hyperbolic tetrahedra, on which the compact closure constraint would generate well-defined gauge transformations. We believe that \cite{Haggard:2014xoa} offers a good starting point towards such a goal.

\item We would need to investigate whether all our framework is generalizable beyond the tetrahedron to hyperbolic polyhedra with arbitrary number of faces, and how it would fit with the deformed loop gravity phase space structure and its geometrical interpretation.

\end{itemize}


We believe that these points would constitute a substantial progress towards understanding the q-deformation of loop quantum gravity as accounting for a cosmological constant in $3+1$-dimensions, thus generalizing the previous work  \cite{HyperbolicPhaseSpace} dealing with discrete hyperbolic geometries in $2+1$-dimensions.

%

\appendix

\section{$\SL(2,\C)$: relating the Cartan and Iwasawa decomposition}
\label{app:relation}
We quickly establish the formulas relating the Cartan and Iwasawa decomposition of $SL(2,\C)$ matrices.

So let $\Lambda$ be in $SL(2,\C)$. There is a unique couple $(B,H)$ in $SH_2(\C) \times SU(2)$ such as:
\begin{equation}
\Lambda = BH
\end{equation}
This is the Cartan decomposition.

Now, there is also a unique couple $(L,H)$ in $SB(2,\C) \times SU(2)$ such as:
\begin{equation}
\Lambda = LH
\end{equation}
This is the Iwasawa decomposition if we choose $L$ with the following restriction: the diagonal must be real and positive.

Now can we relate this two decomposition. Let $\Lambda$ be an $SL(2,\C)$ matrix and $(B,L,H_B,H_L)$ be in $SH_2(\C) \times SB(2,\C) \times SU(2) \times SU(2)$ with:
\begin{equation}
\Lambda = BH_H = LH_L
\end{equation}
First, let's decompose $L$ into the Cartan decomposition:
\begin{equation}
L = BH
\end{equation}
We look for solution of the form:
\begin{equation}
H = \alpha \left(L + (L^{-1})^\dagger\right)
\end{equation}
We consider the constraint $H H^\dagger = 1$:
\begin{eqnarray*}
\alpha \alpha^* \left(L + (L^{-1})^\dagger\right) \left(L^\dagger + L^{-1}\right) &=& 1 \\
\Leftrightarrow \alpha \alpha^* \left( L L^\dagger + 1 + 1 + (L^{-1})^\dagger L^{-1} \right) &=& 1
\end{eqnarray*}
We have: $(L^{-1})^\dagger L^{-1} + L L^\dagger \propto 1$ because $L L^\dagger$ is hermitian. So:
\begin{equation}
\left(2 + \mathrm{Tr}\left(L^\dagger L\right)\right) | \alpha |^2 = 1
\end{equation}
For $L$ with positive diagonal, $H$ has a real determinant if and only if $\alpha$ is real. We can similarly show that $\alpha$ must be positive in order for the determinant to be positive. Those conditions are requiredd as $H$ has determinant $1$. So:
\begin{equation}
H = \frac{L + (L^{-1})^\dagger}{\sqrt{2 + \mathrm{Tr} LL^\dagger}}
\end{equation}
Now we can write the $B$ that goes with it:
\begin{equation}
B = \frac{1+ L L^\dagger}{\sqrt{2 + \mathrm{Tr} LL^\dagger}}
\end{equation}
which is clearly hermitian. If $BH = L$ then we found the decomposition that we already know to be unique. We find:
\begin{equation}
BH = \frac{2L + (L^{-1})^\dagger LL^\dagger L}{2 + \mathrm{Tr} LL^\dagger}
\end{equation}
By explicitly writing $L$ as:
\begin{equation}
L = \begin{pmatrix}
\lambda & 0 \\
z & \lambda^{-1}
\end{pmatrix}
\end{equation}
we find:
\begin{equation}
(L^{-1})^\dagger + LL^\dagger L = \mathrm{Tr} (LL^\dagger) L
\end{equation}
And so:
\begin{equation}
BH = L
\end{equation}

Now, getting back to our initial decomposition, we have:
\begin{equation}
\Lambda = LH_L = BH H_L
\end{equation}
By uniqueness of the decompositions, we can therefore relate the two decompositions:
\begin{eqnarray}
B &=& \frac{1 + LL^\dagger}{\sqrt{2 + \mathrm{Tr} LL^\dagger}} \\
H_B &=& \frac{L + (L^{-1})^\dagger}{\sqrt{2 + \mathrm{Tr} LL^\dagger}} H_L
\end{eqnarray}

\section{Uniqueness of the solution of the non-compact closure constraints}
\label{app:derivation}

We want to solve:
\begin{equation}
\ell^{-1} L_{dab} \ell^{-1} L_{bcd}^{-1} = \ell L_{abc} \ell L_{abc}
\end{equation}
for $\ell$. We write:
\begin{eqnarray}
\ell &=& \begin{pmatrix}
\lambda & 0 \\
z & \lambda^{-1}
\end{pmatrix} \\
L_{abc} &=& \begin{pmatrix}
\mu_{abc} & 0 \\
\alpha_{abc} & \mu_{abc}^{-1}
\end{pmatrix} \\
L_{dab} &=& \begin{pmatrix}
\mu_{dab} & 0 \\
\alpha_{dab} & \mu_{dab}^{-1}
\end{pmatrix} \\
L_{bcd} &=& \begin{pmatrix}
\mu_{bcd} & 0 \\
\alpha_{bcd} & \mu_{bcd}^{-1}
\end{pmatrix}
\end{eqnarray}
We can now rewrite all the terms of our equation:
\begin{eqnarray*}
\ell^{-1} L_{dab} \ell^{-1} L_{bcd}^{-1} &=& \begin{pmatrix}
\lambda^{-1} & 0 \\
-z & \lambda
\end{pmatrix} \begin{pmatrix}
\mu_{dab} & 0 \\
\alpha_{dab} & \mu_{dab}^{-1}
\end{pmatrix}
 \begin{pmatrix}
\lambda^{-1} & 0 \\
-z & \lambda
\end{pmatrix} \begin{pmatrix}
\mu_{bcd}^{-1} & 0 \\
-\alpha_{bcd} & \mu_{bcd}
\end{pmatrix} \\
&=& \begin{pmatrix}
\lambda^{-2} \mu_{dab} \mu_{bcd}^{-1} & 0 \\
- \left(\lambda^{-1}\mu_{dab} + \lambda \mu_{dab}^{-1}\right)\mu_{bcd}^{-1} z + \mu_{bcd}^{-1} \alpha_{bad} - \lambda^2 \alpha_{bcd} & \lambda^2 \mu_{dab}^{-1} \mu_{bcd}
\end{pmatrix}
\end{eqnarray*}
and:
\begin{eqnarray*}
\ell L_{abc} \ell L_{abc} &=& \begin{pmatrix}
\lambda & 0 \\
z & \lambda^{-1}
\end{pmatrix} \begin{pmatrix}
\mu_{abc} & 0 \\
\alpha_{abc} & \mu_{abc}^{-1}
\end{pmatrix}
 \begin{pmatrix}
\lambda & 0 \\
z & \lambda^{-1}
\end{pmatrix} \begin{pmatrix}
\mu_{abc} & 0 \\
\alpha_{abc} & \mu_{abc}^{-1}
\end{pmatrix} \\
&=& \begin{pmatrix}
\lambda^2 \mu_{abc}^2 & 0 \\
\lambda\mu_{abc}^2 z + \lambda^{-1} z + \mu_{abc} \alpha_{abc} + \lambda^{-2} \alpha_{abc} & \lambda^{-2} \mu_{abc}^{-2}
\end{pmatrix}
\end{eqnarray*}
The bottom left component gives a linear equation in $z$ which can be solved. We just have to show that the solution to:
\begin{equation}
\lambda^{-2} \mu_{dab} \mu_{bcd}^{-1} = \lambda^2 \mu_{abc}^2
\end{equation}
is unique. This is easy, as we get:
\begin{equation}
\lambda^4 = \frac{\mu_{dab} \mu_{bcd}^{-1}}{\mu_{abc}^2}
\end{equation}
which has only one real positive solution.


\bibliographystyle{bib-style}
\bibliography{qspace}

\end{document}